\documentclass[a4paper,preprint,pra,superscriptaddress,amsmath,amssymb,,floatfixmathrsfs]{revtex4-1}
\usepackage{tabularx}
\usepackage{graphicx}
\usepackage{braket}
\usepackage{dcolumn}
\usepackage{color}

\begin{document}

\title{Principles of enantio-selective excitation in three-wave mixing spectroscopy of chiral molecules}

\author{Monika Leibscher} 
 \affiliation{Theoretische Physik, Universit\"at Kassel, Heinrich-Plett-Stra{\ss}e 40, 34132 Kassel}
 \author{Thomas F. Giesen}
 \affiliation{Experimentalphysik, Universit\"at Kassel, Heinrich-Plett-Stra{\ss}e 40, 34132 Kassel}
 \author{Christiane P. Koch}
 \affiliation{Theoretische Physik, Universit\"at Kassel, Heinrich-Plett-Stra{\ss}e 40, 34132 Kassel}
\date{\today}

\begin{abstract}
  Three-wave mixing spectroscopy 
  of chiral molecules, which exist in left-handed and right-handed conformations, allows for enantio-selective population transfer despite random orientation of the molecules. This is based on constructive interference of 
the three-photon pathways  for one enantiomer and destructive one for the other. We prove here that three mutually orthogonal polarization directions are required to this end. Two different dynamical regimes exist to realize enantio-selective population transfer, and we show that they correspond to different phase conditions in the three-wave mixing. We find the excitation scheme used in current rotational three-wave mixing experiments of chiral molecules with $C_1$ symmetry to be close to optimal and discuss prospects for ro-vibrational three-wave mixing experiments of axially chiral molecules. Our comprehensive study allows us to clarify earlier misconceptions in the literature.
\end{abstract}

\maketitle

\section{Introduction}

Chiral molecules exist in left-handed and right-handed conformations referred to as enantiomers. An ensemble of chiral molecules typically consists of  a racemat, i.e., a statistical mixture containing 50\% of each enantiomer. In the gas phase, the means for enantiomer separation, conversion and purification are essentially limited to electromagnetic fields. Their use to this end has been discussed theoretically for about two decades~\cite{FujimuraAngewandte00,UmedaJACS00,HokiChemPhys01,
GonzalezPCCP05,ShapiroPRL00,KralPRL01,KralPRL03,FrishmanJCP03,
ThanopulosJCP03,GerbasiJCP04,JacobJCP12,HirotaPJA12,LehmannJCP18,Vitanov19}. Even simply separating the two enantiomers out of a racemat is already a formidable task because of their identical spectra, except for a tiny energy difference  due to the weak interaction~\cite{QuackCPL86,FabriCPC15}. A possible handle for separation arises for molecules with three non-vanishing dipole moment components,  ${\mu_a}$, ${\mu_b}$, $ {\mu_c}$, in the molecule-fixed coordinates, since one component switches sign upon change of enantiomer such that the enantiomers can exhibit a different interaction with electromagnetic radiation. Enantiomer separation exploiting this difference to achieve enantio-selective population of certain molecular states can make use of the  electronic~\cite{FujimuraAngewandte00,UmedaJACS00,HokiChemPhys01,
GonzalezPCCP05}, vibrational \cite{KralPRL01,KralPRL03,JacobJCP12}, or rotational \cite{HirotaPJA12,PattersonNature13,PattersonPCCP14,ShubertAngewandte14,LobsigerJPCL15,
EibenbergerPRL17,PerezAngewandte17,DomingosAnnuRevPhysChem18,LehmannJCP18} degrees of freedom of the molecule. 

First theoretical proposals for enantio-selective excitation assumed the molecules to be oriented in space~\cite{FujimuraAngewandte00,UmedaJACS00,HokiChemPhys01,
GonzalezPCCP05,ShapiroPRL00}. However, this is not a fundamental requirement since the enantiomer-specific electric-dipole interaction survives 
orientational averaging~\cite{ShapiroPRL03Erratum,FrishmanJCP03,ThanopulosJCP03}. In particular, three orthogonal, linearly polarized electromagnetic fields result in enantio-selective excitation of rotational~\cite{LehmannJCP18} or  ro-vibrational  states~\cite{ShapiroPRL03Erratum,FrishmanJCP03,ThanopulosJCP03}.
Experimentally, enantio-selective excitation was demonstrated with three-wave mixing microwave spectroscopy  of rotational states, using resonant excitation of states that are connected by the $a-$type, $b-$type, and $c-$type component of the dipole moment~\cite{PattersonNature13,PattersonPCCP14,ShubertAngewandte14,LobsigerJPCL15,
EibenbergerPRL17,PerezAngewandte17,DomingosAnnuRevPhysChem18}. This is possible for chiral molecules with C$_1$-symmetry, i.e., molecules without a rotation axis, which have three non-vanishing components of their permanent dipole moment. In contrast, axially chiral molecules possess a rotational axis and therefore have only one non-vanishing component of their permanent dipole moment. However, transition dipole moments exist for all three projections such that excitation of ro-vibrational  states may give rise to enantio-selectivity~\cite{ThanopulosJCP03,JacobJCP12}. Similarly, induced dipole moments that come into play when the light is detuned far off resonance with any molecular transition may serve the same purpose
\cite{YachmenevPRL16,GershnabelPRL18,TutunnikovJPCL18}.

Cyclic coupling of three molecular states is a common feature of  enantio-selective excitation schemes for randomly oriented chiral molecules using the rotational or rovibrational structure~\cite{KralPRL01,KralPRL03,FrishmanJCP03,
ThanopulosJCP03,GerbasiJCP04,JacobJCP12,HirotaPJA12,Vitanov19,LehmannJCP18,LiPRA18}. 
This is rationalized by the fact that three-wave mixing spectroscopy, as a purely electric-dipole-based technique, requires a vectorial observable for enantio-selectivity~\cite{OrdonezPRA18}. Whether further fundamental requirements have to be met and whether there  is an optimal way to implement the cyclic coupling are currently open questions. For example, recent microwave experiments~\cite{PattersonNature13,PattersonPCCP14,ShubertAngewandte14,
EibenbergerPRL17,PerezAngewandte17} have used electromagnetic fields with three perpendicular polarizations. In contrast, some of the theoretical proposals employ electromagnetic fields of a single~\cite{ShapiroPRL00,JacobJCP12,HirotaPJA12}, two~\cite{JacobJCP12}, or three~\cite{LehmannJCP18,LiPRA18} polarization directions. Moreover, adiabatic \cite{KralPRL01,KralPRL03,ThanopulosJCP03,GerbasiJCP04,JacobJCP12} as well as sudden, non-adiabatic population transfer \cite{ShapiroPRL00, PattersonNature13,PattersonPCCP14,ShubertAngewandte14,
EibenbergerPRL17,PerezAngewandte17} has been suggested for enantio-separation, often in combination with coherent control or optimal control theory, 
and also shortcuts to adiabaticity~\cite{Vitanov19}
may be utilized.

The aim of this paper is to clarify which requirements are essential for enantio-selective excitation of bound molecular states. We answer the question of how many polarization directions are needed in Sec.~\ref{sec:cond} by making use of the  rotational symmetry of chiral molecules. Employing the simplest model for cyclic coupling in Sec.~\ref{sec:3level}, we identify the specific conditions for enantio-selectivity in the adiabatic and non-adiabatic regimes. We furthermore show how these two regimes for  enantio-separation can be applied to real molecular systems in Sec.~\ref{sec:real}, considering both  purely rotational excitation of C$_1$-symmetric molecules, i.e., molecules where all three components of the permanent molecular dipole moment are non-zero, and  excitation of rotational  and vibrational  states which can be realized also in chiral molecules with C$_1$ as well as C$_2$-symmetry. We summarize our findings in Sec.~\ref{sec:concl}.

\section{Conditions for enantio-selective excitation}
\label{sec:cond}
We assume the chiral molecules to be rigid enough to model them as  an asymmetric top with Hamiltonian
\begin{equation}
  \hat H_{rot} =  A {\hat J_a}^2 + B {\hat J_b}^2 + C {\hat J_c}^2,
  \label{hrot}
\end{equation}
where $\hat J_a$, $\hat J_b$ and $\hat J_c$ are the angular momentum operators with respect to the principle molecular axes,
and $A > B > C$ are the rotational constants. Different vibrational states can simply be included by means of a tensor product, provided the ro-vibrational coupling is negligible. In other words, accounting for the vibrational dynamics of the molecule does not change the conclusions drawn from the rotational structure. 

For $B=C$ or $A=B$, the molecule becomes a prolate, respectively oblate, symmetric top with 
eigenfunctions $|J, K_a, M \rangle$ or $|J, K_c, M \rangle$. The symmetric top wavefunctions are determined by the rotational quantum number $J$ ($J=0, 1,2,\ldots$) and the quantum numbers $M$ and $K$ ($M,K=-J, -J+1,\ldots,J$) which describe the rotation with respect to a space-fixed axis and a molecule-fixed axis, respectively. The eigenfunctions of the asymmetric top are expressed as superpositions of symmetric top eigenstates,
\begin{equation}
   | J, \tau, M \rangle = \sum_K c_K^{J,M} (\tau) |J,K,M \rangle,  
   \label{asym_top}
\end{equation}
where $K$-states with the same $J$ and $M$ are mixed. Here, $\tau=1,2,\ldots,2J+1$ counts the asymmetric top eigenstates with same $J$ and $M$.
The interaction of the molecule with an electro-magnetic field is described in the dipole-approximation by
\begin{equation}
  \hat H_{int} = - \hat{\vec \mu} \cdot {\vec E(t)},
\end{equation}
where $\hat{\vec \mu}^{\,T} = (\hat\mu_x, \hat\mu_y, \hat\mu_z )$ is the molecular dipole moment in spaced-fixed coordinates. Transformation to the molecule-fixed frame leads to
\cite{JacobJCP12,Zare88}
\begin{eqnarray}
  \hat H_{int}^z &=& - \hat\mu_z E_z (t) \nonumber \\
                 &=& - E_z (t) \left [ \hat\mu_a D_{00}^1 - \frac{\hat\mu_b}{\sqrt{2} } \left ( D_{01}^1 - D_{0-1}^1 \right)
                     +  i \frac{\hat\mu_c}{\sqrt{2}} \left ( D_{01}^1 + D_{0-1}^1 \right) \right]
                     \label{Hintz}
\end{eqnarray}
for an electric field which is linearly polarized along the space-fixed $z$-axis. $D_{MK}^J$ denote the elements of the Wigner $D$-matrix, and $\hat\mu_i$ with $i=a,b,c$ are the components of the dipole moment in the molecule-fixed coordinate system. 
For electric fields polarized linearly along the space-fixed $x$- and $y$-axes, the interaction Hamiltonian reads \cite{JacobJCP12,Zare88}
\begin{eqnarray}
  \hat H_{int}^x &=& - \hat\mu_x E_x (t)  \nonumber \\
                 &=& - E_x (t) \left [ \frac{\hat\mu_a}{\sqrt{2}}  \left( D_{-10}^1  - D_{10}^1 \right)  + \frac{\hat\mu_b}{ 2} \left ( D_{11}^1 - D_{1-1}^1  - D_{-11}^1 + D_{-1-1}^1\right) \right. \nonumber \\
                 &  &  \left. -  i \frac{\hat\mu_c}{2} \left ( D_{11}^1 + D_{1-1}^1  - D_{-11}^1 - D_{-1-1}^1\right) \right]
                      \label{Hintx}                 
\end{eqnarray}                   
and 
\begin{eqnarray}
  \hat H_{int}^y &=& - \hat\mu_y E_y (t)  \nonumber \\
                 &=& - E_y (t) \left [ -i \frac{\hat\mu_a}{\sqrt{2}}  \left( D_{-10}^1  + D_{10}^1 \right)  +  i \frac{\hat\mu_b}{ 2} \left ( D_{11}^1 - D_{1-1}^1  + D_{-11}^1 - D_{-1-1}^1\right) \right. \nonumber \\
                 &  &  \left. +   \frac{\hat\mu_c}{2} \left ( D_{11}^1 + D_{1-1}^1  + D_{-11}^1 + D_{-1-1}^1\right) \right]\,,     
                      \label{Hinty}             
\end{eqnarray}
respectively.
The  elements of the Wigner $D$-matrix, $D_{MK}^J$, determine the selection rules. In the symmetric top basis and using Wigner $3j$-symbols \cite{Zare88},
\begin{eqnarray}
  \langle J'', K'', M'' | D^1_{MK} | J', K', M' \rangle &=& \sqrt{2 J'' +1} \sqrt{2 J' +1} (-1)^{M''+K''}  \nonumber \\
                                                        &\times&  \left(    \begin{array}{ccc}  J' & 1& J'' \\ M' & M & -M'' \end{array} \right)
                                                                                                                        \left( \begin{array}{ccc} J' & 1 & J'' \\ K' & K & -K'' \end{array} \right)\,. \label{w3j}
\end{eqnarray}

\begin{figure}[tbp]
  \centering
  \includegraphics[width=0.5\linewidth]{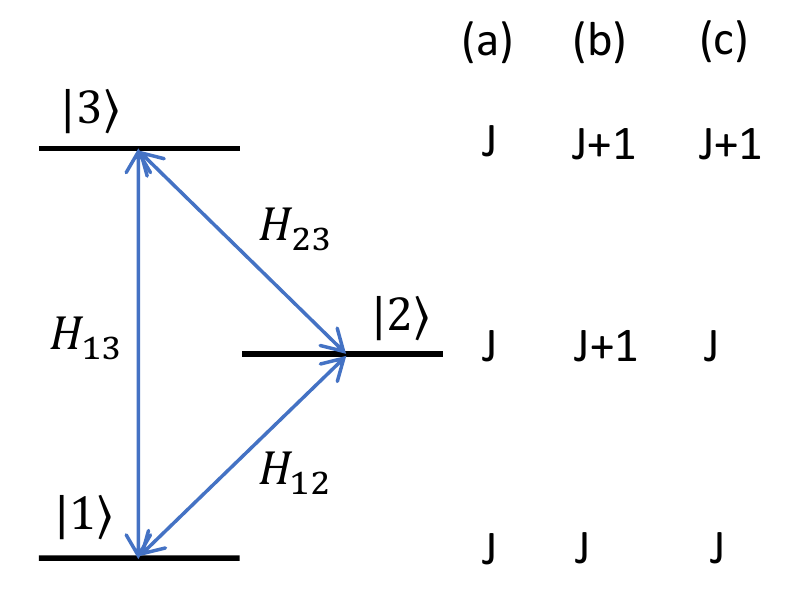}
  \caption{Scheme for cyclic population transfer between three rotational states. The possible combinations of J-states for such three-level cycles are denoted by (a), (b), and (c).}
  \label{scheme_3level}
\end{figure}
To investigate enantio-selective population transfer between rotational states, we consider cyclic coupling of rotational states by three electric fields with frequencies resonant to the three transitions, as shown in Fig.~\ref{scheme_3level}. For electric dipole interaction $\Delta J = 0$ or $\Delta J =\pm 1$. Cyclic three state coupling can thus occur between either three states with same $J$ or two states with same $J$ and one state with $J\pm1$. It is important to note that each of the levels is $(2J+1)$-fold degenerate. 
One should further recall that, for interaction with a field linearly polarized along the $z$-direction, $\Delta M = 0$, whereas $\Delta M = \pm1 $ for interaction with fields linearly polarized along the $x$- and $y$-axes. Moreover, the dipole moment $\hat\mu_a$ corresponds to a transition with $\Delta K = 0$, while $\hat\mu_b$ and $\hat\mu_c$ are responsible for transitions between states with $ \Delta K = \pm 1$.
 
An early proposal for cyclic coupling between three rotational  states suggested use of three electromagnetic fields with the same linear polarization, i.e., ${\vec E(t)} = E_z(t)\,{\vec e_z}$~\cite{ShapiroPRL00}. However, such a cycle cannot exist for $M=0$, since in that case transitions
 with $\Delta J = 0$ are forbidden. Cycles with $M \neq 0$ do exist, but enantio-selectivity was proven to vanish after orientation averaging~\cite{ShapiroPRL03Erratum,FrishmanJCP03}, i.e., after averaging over all $M= \pm |M|$. Cycles with two different polarization directions involve combinations such as $E_z, E_x, E_x$ or $E_z, E_y, E_y$ which were proposed in Ref.~\cite{JacobJCP12}. However, we show in Appendix~\ref{sec:symmreq} that, as a consequence of the permutation symmetry of the Wigner $3j$-symbols, such loops are not enantio-selective. Our group theoretical arguments invalidate
the prediction of a small enantio-selective effect for excitation by pulses with two or even a single polarization direction~\cite{JacobJCP12}. 
Instead, we find in Appendix~\ref{subsec:proof:pol} and in accordance with \cite{LehmannJCP18} that  enantio-selective cyclic three-state coupling requires a combination of fields with three orthogonal polarization directions. 
The symmetry of the rotational states of an asymmetric top implies that such a cyclic three level system always consists of transitions involving all three dipole moment components, $\hat\mu_a$, $\hat\mu_b$ and $\hat\mu_c$ (see Appendix~\ref{subsec:proof:moments}). 
In summary, we give a group-theoretical proof for the statement that three mutual orthogonal polarization directions are necessary for enantio-selective three-wave mixing spectroscopy. This is in accordance with Ref. \cite{LehmannJCP18} and has been at the core of the microwave three-wave mixing experiments of Refs. \cite{PattersonNature13,PattersonPCCP14,ShubertAngewandte14,LobsigerJPCL15, EibenbergerPRL17,PerezAngewandte17,DomingosAnnuRevPhysChem18}
If this condition is fulfilled, enantio-selective cyclic coupling survives orientational averaging and can thus be realized in real molecular systems with degenerate $M$-states. In the following, we use a three-level system with cyclic coupling to investigate various regimes for enantio-selective 
population transfer.

\section{Three-level model for enantio-selective excitation}
\label{sec:3level}

\subsection{Model and field-dressed spectrum}
\label{subsec:model}

Enantio-selective cyclic coupling schemes for randomly oriented chiral molecules involve necessarily degenerate $M$-levels such that the smallest model (with $M=0$ and $M=\pm 1$) consists of four levels. However, we show in Appendix~\ref{subsec:proof:pol} that any such coupling scheme with degenerate $M$-levels can be decomposed into two or more equivalent three-level systems.  
The most elementary model to describe enantio-selective excitation of bound molecular states is thus given by a three-level system. The bare Hamiltonian,
\begin{eqnarray}
 \hat H_0 = \left( \begin{array}{ccc}
           E_1 & 0 & 0 \\
           0     & E_2 & 0 \\
           0& 0& E_3 
           \end{array} \right )\,,
\end{eqnarray}
contains the energies $E_n$ of the rotational or ro-vibrational  states of the  chiral molecule and is identical for both enantiomers. We consider electric-dipole interaction with three linearly polarized fields. The parameters of the fields -- wavelengths, pulse duration and intensity -- are chosen such that  each field resonantly drives only a single transition. The interaction Hamiltonian then reads 
\begin{eqnarray}
 \hat H_{int}^{(\pm)}(t) = \left( \begin{array}{ccc}
           0 & H_{12}^{(\pm)}(t) & H_{13}^{(\pm)}(t) \\
           H_{12}^{(\pm)\ast}(t)    & 0 & H_{23}^{(\pm)}(t) \\
           H_{13}^{(\pm)\ast}(t)    & H_{23}^{(\pm)\ast}(t)  & 0  
           \end{array} \right )\,,
\end{eqnarray}
where
\begin{equation}
  \hat H_{nm}^{(\pm)}(t) = \langle n | - {\vec \mu^{(\pm)}} \cdot {\vec {E}_{\alpha} } | m \rangle 
\end{equation}
with  ${\vec \mu}^{(\pm)}$ the molecular dipole moment and ${\vec {E}_{\alpha}(t) } = {\vec e}_{\alpha} {\cal E}_{\alpha}(t)$
the electric field with polarization direction ${\vec e}_{\alpha}$ and   \begin{equation}
  {\cal E}_{\alpha}(t)  = \epsilon_{\alpha}(t) \cos (\omega_{\alpha}  t + \phi_{\alpha}).
\end{equation}
Here,  $ \epsilon_{\alpha}(t)$ is the envelope of electric field ${\cal E}_{\alpha}(t)$ ($\alpha=x$, $y$, $z$),  whereas its frequency and phase are denoted by $\omega_{\alpha}$ and $\phi_{\alpha}$. The superscript $(\pm)$ refers to the two enantiomeres.

In the interaction picture, the time-dependent Schr\"odinger equation reads
\begin{equation}
   i  \hbar \frac{\partial}{\partial t}|\psi_I^{(\pm)} (t) \rangle = \hat H_I^{(\pm)}(t)  |\psi_I^{(\pm)} (t) \rangle,
   \label{Schr_in}
\end{equation}
with
\begin{eqnarray}\label{eq:H_I}
 H_{I}^{(\pm)}(t) = \left( \begin{array}{ccc}
           \delta_{12} & {\tilde H}_{12}^{(\pm)}(t) & {\tilde H}_{13}^{(\pm)}(t) \\
           {\tilde H}_{12}^{(\pm)\ast}(t)    & 0 & {\tilde H}_{23}^{(\pm)}(t) \\
           {\tilde H}_{13}^{(\pm)\ast}(t)    & {\tilde H}_{23}^{(\pm)\ast}(t)  & -\delta_{23}
           \end{array} \right ).
\end{eqnarray}
Here, the frequencies of the electric fields are $\omega_{z}=(E_2 -E_1)/\hbar + \delta_{12}$, $\omega_{x}=(E_3 -E_1)/\hbar + \delta_{13}$ and $\omega_{y}=(E_3 -E_2)/\hbar + \delta_{23}$.
Within the rotating wave approximation (RWA),
\begin{eqnarray}
  {\tilde H}_{12}^{(\pm)} &=& - \langle 1 | {\vec \mu}^{(\pm)} \cdot  {\vec e}_{z} | 2 \rangle \, \epsilon_{z}(t) \exp \left ( i \phi_{12} \right ) \exp \left [ i \left ( \delta_{12} + \delta_{23} - \delta_{13} \right ) t \right ] \nonumber \\  
   {\tilde H}_{13}^{(\pm)} &=& -  \langle 1 | {\vec \mu}^{(\pm)} \cdot  {\vec e}_{x} | 3 \rangle \, \epsilon_{x}(t) \exp \left ( i \phi_{13} \right ) \nonumber \\
    {\tilde H}_{23}^{(\pm)} &=& -  \langle 2 | {\vec \mu}^{(\pm)} \cdot  {\vec e}_{y} | 3 \rangle \, \epsilon_{y}(t) \exp \left ( i \phi_{23} \right )\,,
 \end{eqnarray}
where we have renamed $\phi_z=\phi_{12}$, $\phi_x=\phi_{13}$ and $\phi_y=\phi_{23}$. 
Recall that, in the RWA, the transition matrix elements  can be complex, i.e.,
 \begin{equation}
 \langle n | {\vec \mu}^{(\pm)}\cdot  {\vec e}_{\alpha} | m \rangle  = \sigma_{nm}^{(\pm)} | \langle n | {\vec \mu}^{(\pm)} \cdot  {\vec e}_{\alpha} | m \rangle | \, \exp \left ( i \theta_{nm} \right ).
 \end{equation} 
It is then useful to define phases \cite{KralPRL01},
\begin{equation}
  \Phi_{nm} =  \theta_{nm} + \phi_{nm}\,,
  \label{phase_overall}
\end{equation} 
which contain the material phase $  \theta_{nm}$ and the phases of the electric fields $\phi_{nm}$. Note that the absolute values of the transition dipole moments are identical for both enantiomers, i.e. $| \langle n | {\vec \mu}^{(+)} \cdot  {\vec e}_{\alpha} | m \rangle | = | \langle n | {\vec \mu}^{(-)} \cdot  {\vec e}_{\alpha} | m \rangle |$ . The difference between the enantiomeres is expressed by the sign $\sigma_{nm}^{(\pm)}$: For two of the three transitions $\sigma_{nm}^{(-)}=\sigma_{nm}^{(+)}=1$. For the third transitions $\sigma_{nm}^{(-)}=-\sigma_{nm}^{(+)}=-1$.
If the detunings from resonance are chosen such that $\delta_{12} + \delta_{23} - \delta_{13} = 0$,
the Hamiltonian $H_I^{(\pm)}$ becomes time-independent except for the slowly varying envelopes of the fields. 
In the following, the frequencies are chosen such that this condition is fulfilled.
To obtain dimensionless units, we scale all energies with some energy $E_0$. A natural choice is $E_0=E_{rot}$, where $E_{rot} = \hbar^2 B$  is the rotational energy, and the rotational constant $B$ is defined in Eq.(\ref{hrot}). Time is then given in units of
$t_0=\hbar/E_{rot}$ and  frequencies  in units of $1/t_0$.

\begin{figure}[tbp]
 \centering
 \includegraphics[width=16cm]{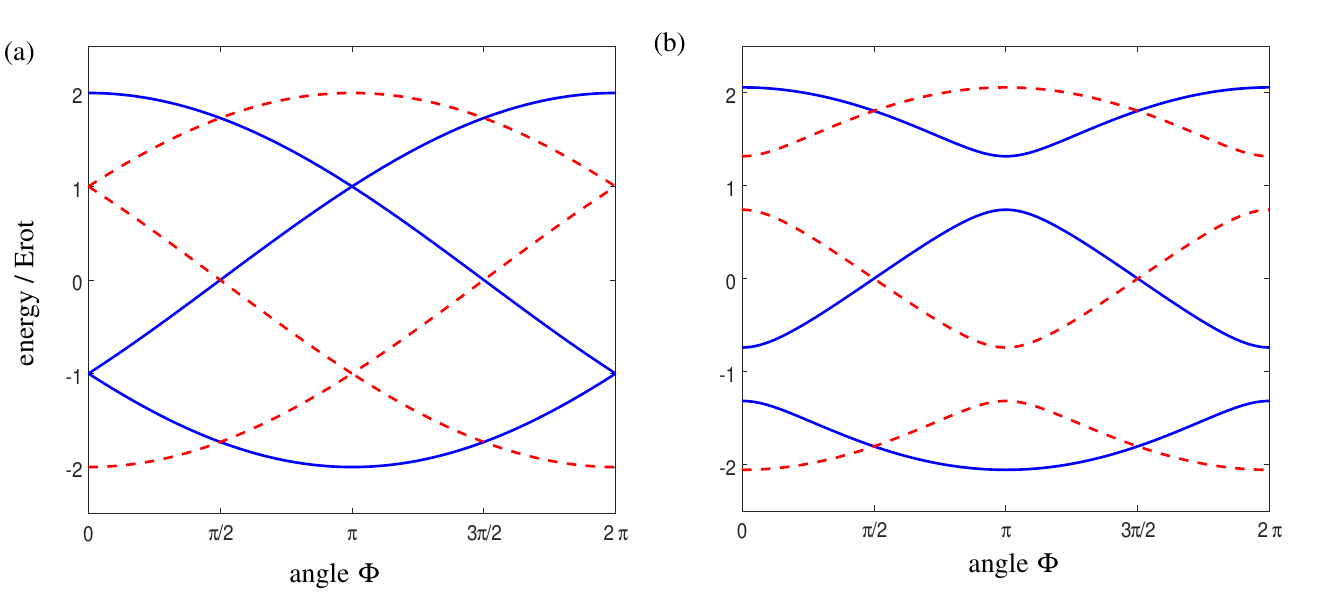}
\caption{Eigenvalues of $H_I^{(\pm)}$ as a function of the overall phase $\Phi$ for (a) zero detuning from resonance, i.e., $\delta_{12}=\delta_{23}=0$, and (b) $\delta_{12}=\delta_{23}=0.5 /t_0$. The two enantiomers are indicated by solid and dashed curves, respectively. All matrix elements $\tilde{H}_{ij}^{(\pm)}$ are assumed to have unit magnitude.  }
\label{spectrum_dressed_states}
\end{figure}
One possibility for enantio-selective population transfer is to use adiabatic following of field-dressed eigenstates~\cite{KralPRL01,KralPRL03,ThanopulosJCP03,GerbasiJCP04,JacobJCP12}. 
In other words, enantio-selectivity of the cyclic coupling should be reflected in the field-dressed eigenvalues. Figure~\ref{spectrum_dressed_states} therefore displays the instantaneous eigenvalues $E_n^{fd,(\pm)}$ obtained by diagonalizing $H_I^{(\pm)}(t)=H_I^{(\pm)}$, Eq.~\eqref{eq:H_I}, for constant field envelope
and resonant (a) as well as near-resonant~(b) excitation, as a function of the overall phase $\Phi= \Phi_{12}+ \Phi_{23}- \Phi_{13}$, cf. Eq.~\eqref{phase_overall}. Note the values   $\pm 1 E_{rot}$  and $\pm 2 E_{rot}$ for the energy of the field-dressed states for $\Phi=0$ result from the assumption that all matrix elements $\tilde{H}_{ij}^{(\pm)}$ have unit magnitude.
 Detuning from resonance lifts the degeneracy of the field-dressed eigenstates within one enantiomer which occurs at  $\Phi=0$, $\pi$, and $2\pi$. Enantio-selectivity in the adiabatic regime requires the two enantiomers (marked by solid and dashed lines, respectively) to have a different field-dressed spectrum.
 As can be seen in 
Fig.~\ref{spectrum_dressed_states}, this is the case for all phases except 
$\Phi=\pi/2$ and $\Phi=3\pi/2$ for both resonant and near-resonant excitation.

Thus, the eigenvalues of the field-dressed Hamiltonian already indicate that enantio-selective population transfer is possible, as is also discussed in Ref.~\cite{HirotaPJA12}, provided specific inauspicious phases are avoided. The difference in the field-dressed spectra suggests to realize enantio-selective population transfer by keeping the system in the same field-dressed state at all times, i.e., to use adiabatic excitation. This regime is inspected below in Sec.~\ref{subsec:adiab}. 
Enantio-selective excitation of rotational  states by non-adiabatic excitation is investigated in Sec.~\ref{subsec:nonadiab}.

\subsection{Enantio-selectivity via simultaneous adiabatic and diabatic passage}
\label{subsec:adiab}

\begin{figure}[tbp]
\centering
\includegraphics[width=14cm]{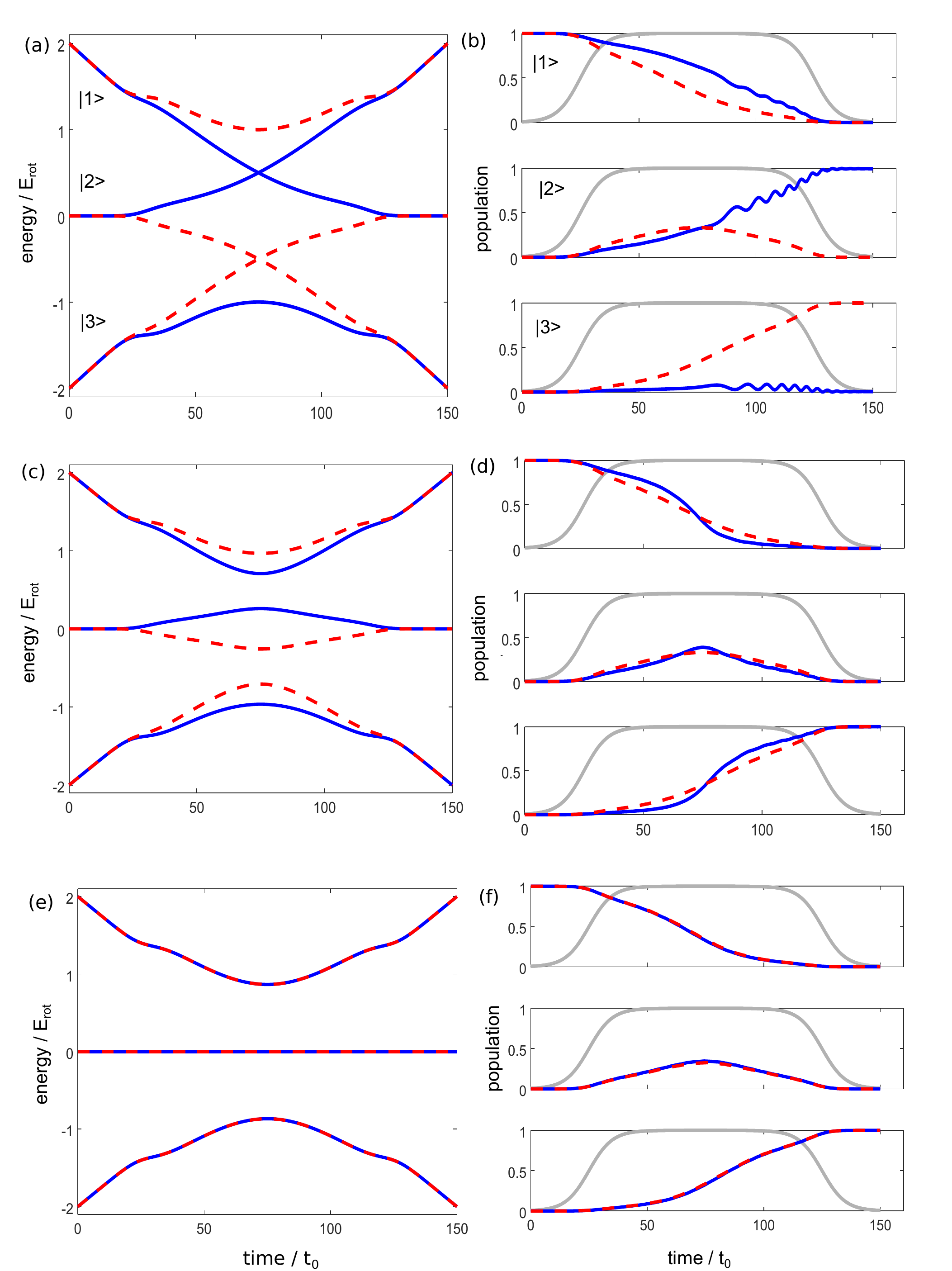}
\caption{Left: Energy of the field-dressed eigenstates during the interaction with linearly chirped pulses. Right: Population of the field-free states $|n\rangle$ (top $n=1$, middle $n=2$, bottom $n=3$) with the envelope of the pulses shown in gray (all three pulses are switched on and off simultaneously). The chirp is defined be the detuning 	$\delta_{12}^0=\delta_{23}^0=2/t_0$, see Eq.(\ref{eq_detuning}).
The phases are $\Phi=0$, $\pi/4$ and $\pi/2$ for the top, middle and bottom parts, respectively. }
\label{fig_adiabatic_spectrum_pop}
\end{figure}
We assume the system to be initially in its ground state,  $|\psi_I^{(\pm)}(t=0) \rangle=|1\rangle$. 
The time evolution of the system occurs in an adiabatic way if the state remains in the same field-dressed eigenstate  $|\psi_I^{(\pm)}(t) \rangle$. Adiabatic population transfer between two field-free molecular states can be realized for example using linearly chirped pulses which induce rapid adiabatic passage~\cite{VitanovAnuRevPhysChem01}. An application of rapid adiabatic passage to enantio-selective excitation is shown in Fig.~\ref{fig_adiabatic_spectrum_pop} where the solid (dashed) line presents enantiomer $(+)$ ($(-)$).
The left panels show the energies of the field-dressed eigenstates for the two enantiomers, while the populations of the field-free states are depicted on the right.  All three electric fields are switched on and off simultaneously; the envelopes of the fields are indicated by gray lines. The pulses are chirped such that
\begin{equation}
     \delta_{nm}(t) = -  \delta_{nm}^0 + \frac{ 2 \delta_{nm}^0 }{\Delta t} t
     \label{eq_detuning} 
\end{equation}
for $nm=12$ or $nm=23$ and $\Delta t$ the interaction time. Note 
that at $t=0$ the highest energy field-dressed state corresponds to the field-free state $|1 \rangle$.

Figure~\ref{fig_adiabatic_spectrum_pop}(a) shows the energies of the field-dressed eigenstates for $\Phi=0$. If the system is initially in the field-free ground state $|1\rangle$, the $(-)$ enantiomer (dashed lines) evolves adiabatically along the highest energy field-dressed eigenstate and is transferred, after the interaction, to the field-free state $|3 \rangle$, as can indeed be seen in Fig.~\ref{fig_adiabatic_spectrum_pop}(b). For enantiomer $(+)$ (solid lines), the two highest field-dressed states cross at $t = \Delta t/2$, where the detuning is zero. Therefore, enantiomer $(+)$ evolves diabatically and is transfered to the field-free state $|2 \rangle$. Thus, for $\Phi=0$, the interaction leads to enantio-selective population transfer.

Figure~\ref{fig_adiabatic_spectrum_pop}(c) and (d) show the field-dressed eigenstates and the population of the field-free states for $\Phi= \pi/4$. 
Here, the field-dressed states of the two enantiomers have different spectra, but there is no degeneracy between different field-dressed states. As a consequence, both enantiomers evolve adiabatically along the highest field-dressed state and are finally  transferred to the same field-free state $| 3 \rangle$. The slight difference in the population during the interaction is a result of the different field-dressed spectra. For $\Phi=\pi/2$, shown in Fig.~\ref{fig_adiabatic_spectrum_pop}(e) and (f),  the population difference vanishes, as expected from the discussion in Sec.~\ref{subsec:model} above, because the field-dressed states of the two enantiomers have identical  spectrum. 

In general, we find that an exact crossing of the field-dressed states occurs only if $\Phi=0$, the detunings are zero and $ |{\tilde H}_{12}^{(\pm)}| = |{\tilde H}_{13}^{(\pm)}| = |{\tilde H}_{23}^{(\pm)}|$. For other sets of parameters, small avoided crossings lead to selective but incomplete population transfer. Maximum enantio-selectivity is only obtained when one enantiomer undergoes adiabatic evolution whereas the other one simultaneously evolves diabatically, i.e., for $\Phi=0$, $\pi$, and $2\pi$.
It also 
requires the three electric fields to be present at the same time. This can be achieved if the fields are turned on and off simultaneously, as shown here.

The application of rapid adiabatic passage to enantio-selective excitation has been proposed in Refs.~\cite{KralPRL01,ThanopulosJCP03}, where time-delayed but overlapping
sequences of three pulses are employed to achieve cyclic population transfer.
Such a sequential excitation of the three transitions opens more possibilities to optimize the driving fields, and coherent control has indeed been used to find pulse parameters for which small avoided crossings between field dressed states lead to selective population transfer \cite{KralPRL01,ThanopulosJCP03}. 
However, the resulting parameters, i.e., the phases of the laser pulses, are specific to the pulse sequence. In contrast, with our more elementary model for cyclic population transfer, we can directly relate the phase condition to the existence of an exact crossing between the field-dressed states, 
as shown above, and are thus able to directly identify the interference 
mechanism enabling cyclic population transfer.

\subsection{Enantio-selectivity via Rabi oscillations in non-adiabatic excitation}
\label{subsec:nonadiab}

\begin{figure}[tbp]
\includegraphics[width=14cm]{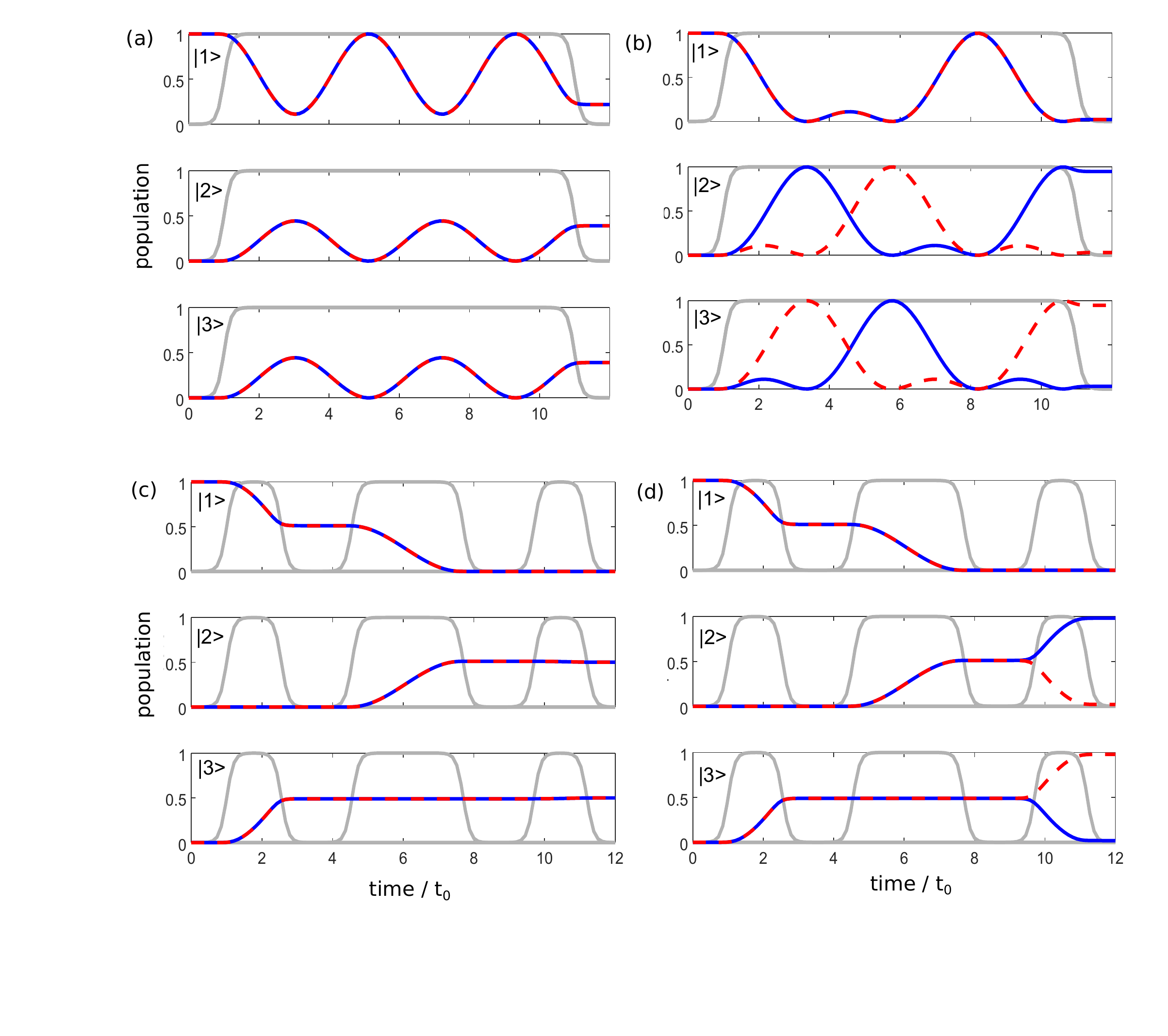}
\caption{Population of the field-free states $|n\rangle$ (small top $n=1$, middle $n=2$, bottom $n=3$ panels) for simultaneous (top part) and sequential (bottom part) resonant excitation with  $\Phi=0$ (left) and $\Phi=\pi/2$ (right).
The two enantiomers are indicated by solid and dashed curves, 
the envelope of the pulses are indicated in gray. 
}
\label{non_adiabatic_pop}
\end{figure}
We assume again  the initial  state to be $|\psi_I^{(\pm)}(t=0)\rangle=|1\rangle$. Without detuning from resonance, excitation with three pulses leads to the creation of a wavepacket consisting of all field-dressed states which results in Rabi oscillations.
The corresponding population of the field-free states $|n\rangle$ can be seen in Fig. \ref{non_adiabatic_pop}.
In the top part of Fig. \ref{non_adiabatic_pop}, 
all three pulses are switched on and off simultaneously. For $\Phi=0$, one of the Rabi frequencies is equal to zero since two of the field-dressed eigenstates are degenerate, as it can be seen in Fig. \ref{spectrum_dressed_states}(a). As a result, the three states cannot be populated selectively, cf. Fig. \ref{non_adiabatic_pop}(a). For $\Phi=\pi/2$, cf. Fig. \ref{non_adiabatic_pop}(b), occurrence of three different Rabi frequencies leads to enantio-selective population of the two exited states, $|2\rangle$ and $|3\rangle$. Remarkably, we find that enantio-selective population transfer can be achieved for any phase $\Phi \neq 0$. For $\Phi \rightarrow 0$, however, the time required for separation approaches infinity.

Instead of applying the three pulses simultaneously, they can also be switched on and off one after the other, as shown  in the bottom part of Fig.~\ref{non_adiabatic_pop}. Here, the first pulse is a $\pi/2$-pulse which creates a 50/50-superposition between the states $|1\rangle$ and $|3\rangle$. The second pulse, a $\pi$-pulse, leads to population transfer from $|1\rangle$ to $|2\rangle$. Finally, the third pulse connects states  $|2\rangle$ and $|3\rangle$, where
constructive versus destructive interference causes the enantio-selective excitation of states $|2\rangle$ and $|3\rangle$, respectively.
This is essentially the scheme that has been realized in the microwave experiments reported in Refs.~\cite{PattersonNature13,PattersonPCCP14,ShubertAngewandte14,
EibenbergerPRL17,PerezAngewandte17}.
Note that any possible $\pi/2$-$\pi$-$\pi/2$-pulse sequence starting with the initially populated state can be used for enantio-selective excitation. The order of the pulses determines for a given enantiomer which final state is populated. Applying for example the first $\pi/2-$pulse to the transition between states $|1\rangle$ and $|3\rangle$ and the $\pi-$pulse to the transition $|2\rangle \rightarrow |3\rangle$  leads to selective population of the states $|1\rangle$ and $|3\rangle$ at the end of the sequence. This is in contrast to a simultaneous application of all three pulses, where the final states are always $|2\rangle$ and $|3\rangle$, provided that  $|1\rangle$ is the initially populated state. 

\begin{figure}[tbp]
	\includegraphics[width=16cm]{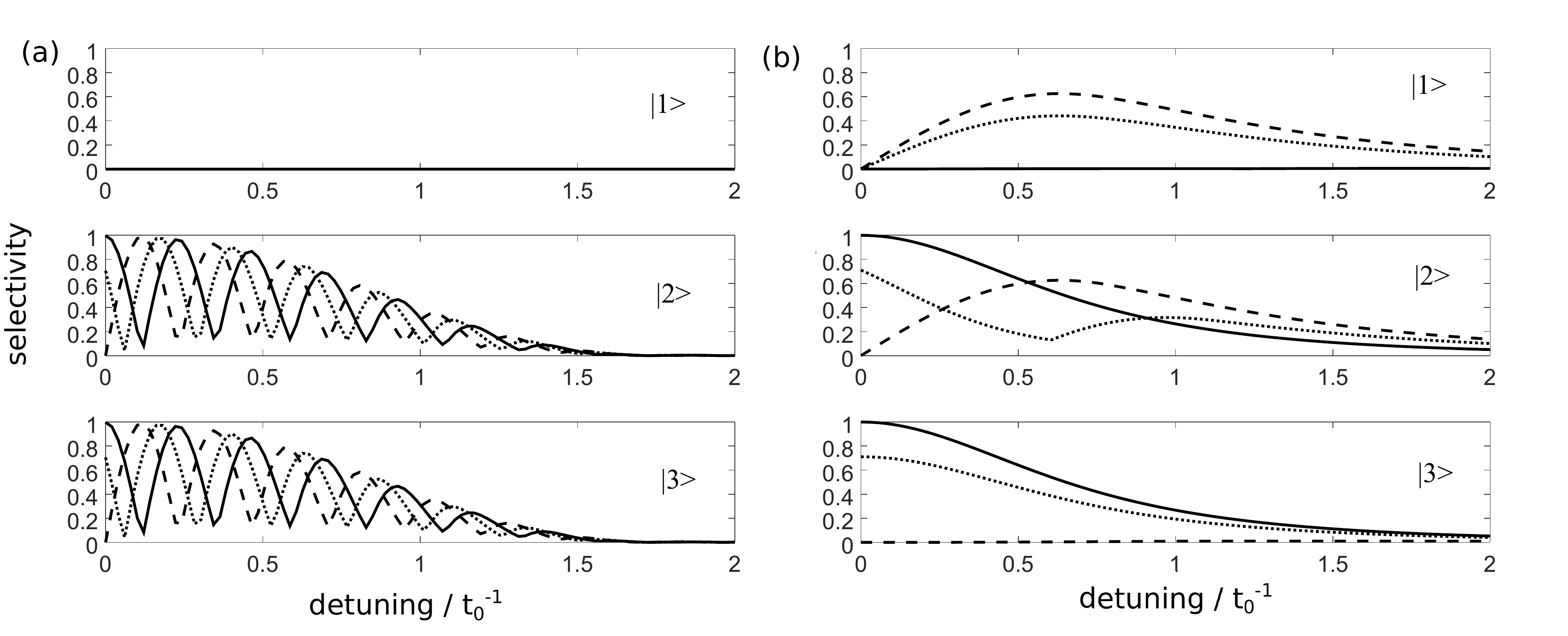}
	\caption{Enantio-selectivity as a function of the detuning from resonance of the field-free states $|n\rangle$ ( top $n=1$, middle $n=2$, bottom $n=3$ panels) for sequential (a) and simultaneous (b) excitation with  $\Phi=0$ (dashed), $\Phi=\pi/4$  (dotted) and  $\Phi=\pi/2$  (solid).}
	\label{detuning_selctivity}
\end{figure}

In the following, we investigate the effects of a (small) detuning from resonance on the enatio-selectivity of non-adiabatic excitation. Figure \ref{detuning_selctivity}, displays the selectivity $|P_n^{(+)} - P_n^{(-)}|$, where $P_n^{(\pm)}$ is the population of state $|n\rangle$, with $n=1,2,3$ of the enatiomer (+) or (-). We consider a constant detuning $\delta_{12}=\delta_{23}$ such that condition $\delta_{12}+\delta_{23}-\delta_{13}=0$ is fulfilled. 
Figure \ref{detuning_selctivity} (a) shows the selectivity for excitation with the same pulse sequence as the one displayed in Fig. \ref{non_adiabatic_pop} (c) and (d). The rapid oscillation of the selectivity as well as the overall decline as the detuning increases can be understood if the excitation process is regarded three two-level systems which are excited successively. In a two-level system, detuning from resonance results in an additional phase of the excited states which leads to a mismatch with the phase condition for selective excitation. For small detunings, this can be compensated for by changing the overall phase $\Phi$. For example, for the pulse sequence considered here, $\Phi=\pi/2$ (solid line) leads to maximal selectivity for $\delta_{12}=\delta_{23}=0$, while for $\delta_{12}=\delta_{23}\approx 0.1$, the phase $\Phi=0$ leads to maximal selectivity. 
With increasing detuning, the selectivity declines for all values of $\Phi$ since detuning from resonance results in incomplete population transfer.

The selectivity as a function of the detuning from resonance for simultaneous excitation
of all three levels can be seen in Fig. \ref{detuning_selctivity} (b). Since all three pulses are applied at the same time and condition $\delta_{12}+\delta_{23}-\delta_{13}=0$ holds, the additional phases for each transition cancel, and rapid oscillation of the selectivity does not occur. For $\Phi=\pi/2$ (solid line) the selectivity declines with increasing detuning due to the incomplete population transfer. Moreover, for non-zero detuning, the degeneracy of the field-dressed states at $\Phi=0$ is lifted (see Fig. \ref{spectrum_dressed_states}). Therefore, excitation with $\Phi=0$ (dashed line) also becomes enatio-selective with increasing detuning.

Enantio-selective excitation with three pulses with pulse areas of $\pi$ and $\pi/2$, respectively has been experimentally demonstrated in microwave three-wave mixing experiments\cite{PattersonNature13,PattersonPCCP14,ShubertAngewandte14,LobsigerJPCL15,
	EibenbergerPRL17,PerezAngewandte17,DomingosAnnuRevPhysChem18} and has also been discussed in \cite{LehmannJCP18}. Here, we have simulated this process in the non-adiabatic regime. This has allowed us to show that the pulses can be applied both  simultaneously or sequentially, without overlap in time, provided the total phase is not equal to zero. Moreover, we have demonstrated that a small detuning from resonance reduces the amount of selectivity, an effect that can be partly overcome by adjusting the phase of the electric fields.

\subsection{Comparison between adiabatic and non-adiabatic excitation}

We have identified two complementary mechanisms for enantio-selective 
population transfer -- adiabatic following of a field-dressed eigenstate for one enantiomer with simultaneous diabatic dynamics for the other enantiomer compared to fully non-adiabatic dynamics of both enantiomers.
While in the simultaneous adiabatic-diabatic regime enantio-selective population transfer requires the overall phase to strictly be $\Phi=0$, 
it can be achieved for nonadiabatic dynamics with all phases $\Phi \neq 0$. Moreover, enantio-selective excitation in the adiabatic-diabatic regime 
requires an exact crossing of the field-dressed energy eigenstates which translates into the requirement that all three pulses overlap in time. This is not necessary, if the rotational states are excited  resonantly in the  
non-adiabatic regime. Here, the pulses can be applied simultaneously as well as sequentially. Sequential excitation comes with the advantage that the parameters for each pulse can be optimized separately. This is important for example when  one of the three dipole matrix elements is much smaller than the other two, as discussed below in Sec.~\ref{sec:real}.
With sequential non-adiabatic excitation it is thus easier to find conditions for optimal selectivity than in the adiabatic-diabatic regime 
where the parameters of all pulses have to be controlled simultaneously.
It should be noted that chirped pulses can also be used sequentially to create a sequence of three pulses with pulse areas of $\pi$ and $2\pi$. The first enantio-selective microwave three-wave mixing experiments have been conducted this way \cite{ShubertAngewandte14}. In that work, chirping the pulses does not provide a realization of the simultaneous adiabatic and diabatic passage proposed in Section III B. Rather, it is a different way to implement the excitation scheme described in Section III C.

\section{Application to real molecules}
\label{sec:real}

We now lift two idealizing assumptions made in Sec. III, namely 
that all states can be addressed individually and that all transition matrix elements have the same magnitude. On other words, we investigate enantio-selectivity for a realistic molecular structure but continue to assume the molecules to be initially in their ground state, i.e., at zero temperature. We discuss the application of both excitation schemes  to purely rotational 
excitation of chiral molecules with $C_1$-symmetry in Sec.~\ref{subsec:rot}
and to ro-vibrational  excitation of axially chiral systems in Sec.~\ref{subsec:rovib}.

\subsection{Application I: Rotational spectroscopy}
\label{subsec:rot} 

\begin{table}
\begin{tabular}{c|ccc|ccc}
     & A [MHz]  & B [MHz]  & C [MHz] & $\mu_a$ [D] & $\mu_b$ [D] & $\mu_c$ [D] \\
     \hline
    menthol  & 1779.8 & 692.63 & 573.34 & 1.3 & 0.1 & 0.8 \\ 
  carvone  &   2237.21 & 656.28 & 579.64 &   2.0 & 3.0  & 0.5 \\    
\end{tabular}
\caption{Rotational constants and magnitude of dipole moments for menthol and carvone. Data are taken from Refs.~\cite{SchmitzFrontiers15} and \cite{MorenoStructural13}.}
\label{rot_constants}
\end{table}
\begin{figure}[tbp]
\centering
\includegraphics[width=8cm]{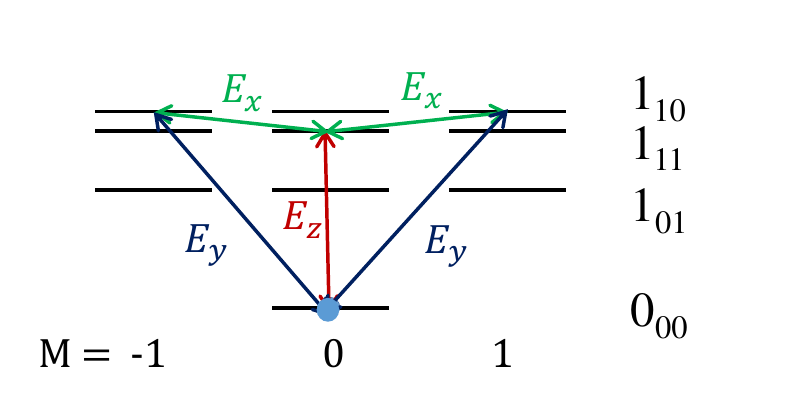}
\caption{Lowest rotational  states of a prolate asymmetric top molecule such as menthol. The arrows indicate the transitions driven by the electric fields $E_x$, $E_y$ and $E_z$. The rotational energy levels are labeled by $J_{K_a K_c}$.}
\label{spectrum_menthone}
\end{figure}
As an example, we consider the chiral molecule menthol with rotational constants and dipole moments summarized in Table~\ref{rot_constants}. 
An enantio-selective cycle can be realized by the combination of microwave pulses depicted in Fig. \ref{spectrum_menthone}, where the transitions driven by $E_x$, $E_y$, and $E_z$ are of $a-$type, $c-$type and $b-$type, respectively. 
Following the convention in microwave spectroscopy, we denote the rotational states by $|J \tau M \rangle = |J_{K_a,K_c,M} \rangle$ in the following, where $K_a$ and $K_c$ are the quantum numbers of a prolate and oblate symmetric top, respectively.
The population of a rotational level $J_{K_a,K_c}$ is averaged over the corresponding $2J+1$ M-states.

\begin{figure}[tbp]
\centering
\includegraphics[width=8cm]{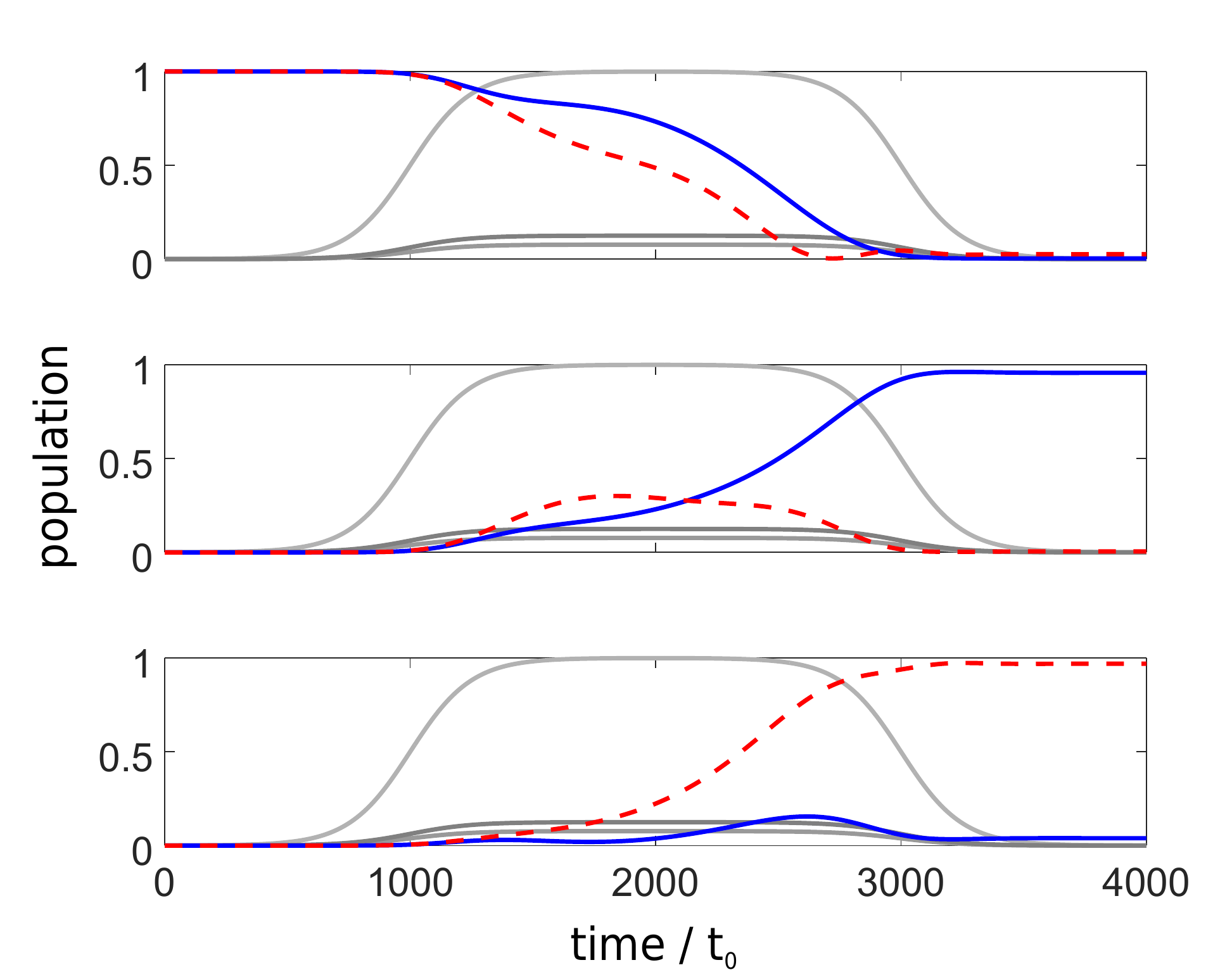}
\caption{Adiabatic population transfer for the rotational levels $0_{00}$ (top),  $1_{10}$ (middle) and $1_{11}$ (bottom) of menthol, averaged over M-states. The two enantiomers are indicated by solid (blue) and dashed (red) lines. The envelope of the (simultaneously applied) pulses is indicated by gray lines. The maximal intensities are $I_z=6.3\,$W/cm$^2$,  $I_y=0.1\,$W/cm$^2$, and $I_x=0.04\,$W/cm$^2$. The detuning, Eq.(\ref{eq_detuning}), is $\delta_{12}^0=\delta_{23}^0=0.01 B$.
Time is given in units of $t_0=\hbar/E_{rot} \approx 1.4$ ns. Here, $\Phi=0$.}
\label{menthone_adiabatic}
\end{figure}
Enantio-selectivity via adiabatic passage for one enantiomer with simultaneous  diabatic passage for the other enantiomer is examined in Fig.~\ref{menthone_adiabatic} for chirped microwave pulses (indicated in gray) and $\Phi=0$. In this regime, as shown in Sec.~\ref{subsec:adiab}, the population transfer is enantio-selective if there is an exact crossing between the energies of two field-dressed states. This translates into the condition 
that all transition matrix elements, which are proportional to the dipole moments, $\mu_a$,  $\mu_b$, $\mu_c$ times the corresponding field amplitudes, have the same magnitude. Since for menthol, $\mu_b \ll \mu_a$, the small $b-$type dipole moment has to be compensated by a higher intensity of the corresponding electric field. We therefore choose the field strengths such that
$\mu_a E_x=\mu_c E_y=\mu_b E_z$.
A selectivity of approximately 95\% is obtained in  Fig.~\ref{menthone_adiabatic} for a pulse duration of $5.6 \mu$s. By further optimization of the parameter - field strengths, pulse duration and detuning from resonance - completely adiabatic excitation and therefore 100\% of separation may be achieved.

\begin{figure}[tbp]
\centering
\includegraphics[width=16cm]{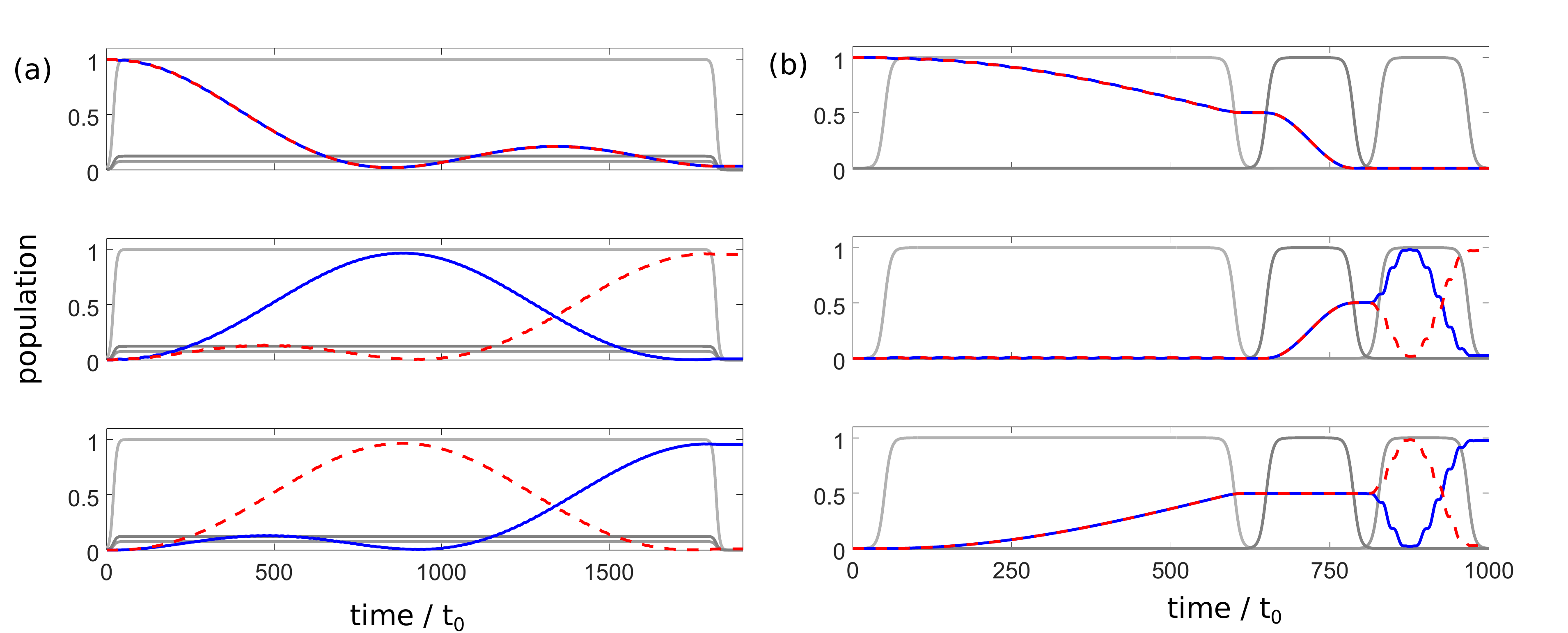}
\caption{Non-adiabatic population transfer for the rotational states $|0_{00}\rangle$ (top),  $|1_{10}\rangle$ (middle) and $|1_{11}\rangle$ (bottom) of menthol.  The frequencies of the three pulses are resonant to the transitions between the three states. The two enantiomers are indicated by solid (blue) and dashed (red) lines and the envelope of the pulses by gray lines. The maximal intensities are (a) $I_z=6.3\,$W/cm$^2$,  $I_y=0.1\,$W/cm$^2$, and $I_x=0.04\,$W/cm$^2$ and (b)  
$I_x=I_y=I_z= 6.3\,$W/cm$^2$. Time is given in units of $t_0=\hbar/E_{rot} \approx 1.4\,$ns. In all cases, $\Phi=\pi/2$.}
\label{fig_menthone_nonadiabatic}
\end{figure}
Non-adiabatic population transfer between rotational states of menthol is shown in Fig.~\ref{fig_menthone_nonadiabatic}. 
Again, maximal selectivity requires the transition matrix elements to have the same magnitude. If the three pulses are switched on and off simultaneously, this is realized, as in Fig.~\ref{menthone_adiabatic}, by adapting the field strengths. The resulting population transfer can be seen in Fig.~\ref{fig_menthone_nonadiabatic}(a), where we obtain nearly 100\% selectivity despite the small level spacing and differences in the pulse intensities.
Non-adiabatic excitation with nearly 100\% selectivity can also be realized by a sequential interaction without any overlap of the three pulses, shown in Fig.~\ref{fig_menthone_nonadiabatic}(b). Here, it is convenient to use the same field strength for all three pulses and compensate for the different magnitudes of the dipole moments by changing the pulse durations.

\begin{figure}[tbp]
\centering
\includegraphics[width=14cm]{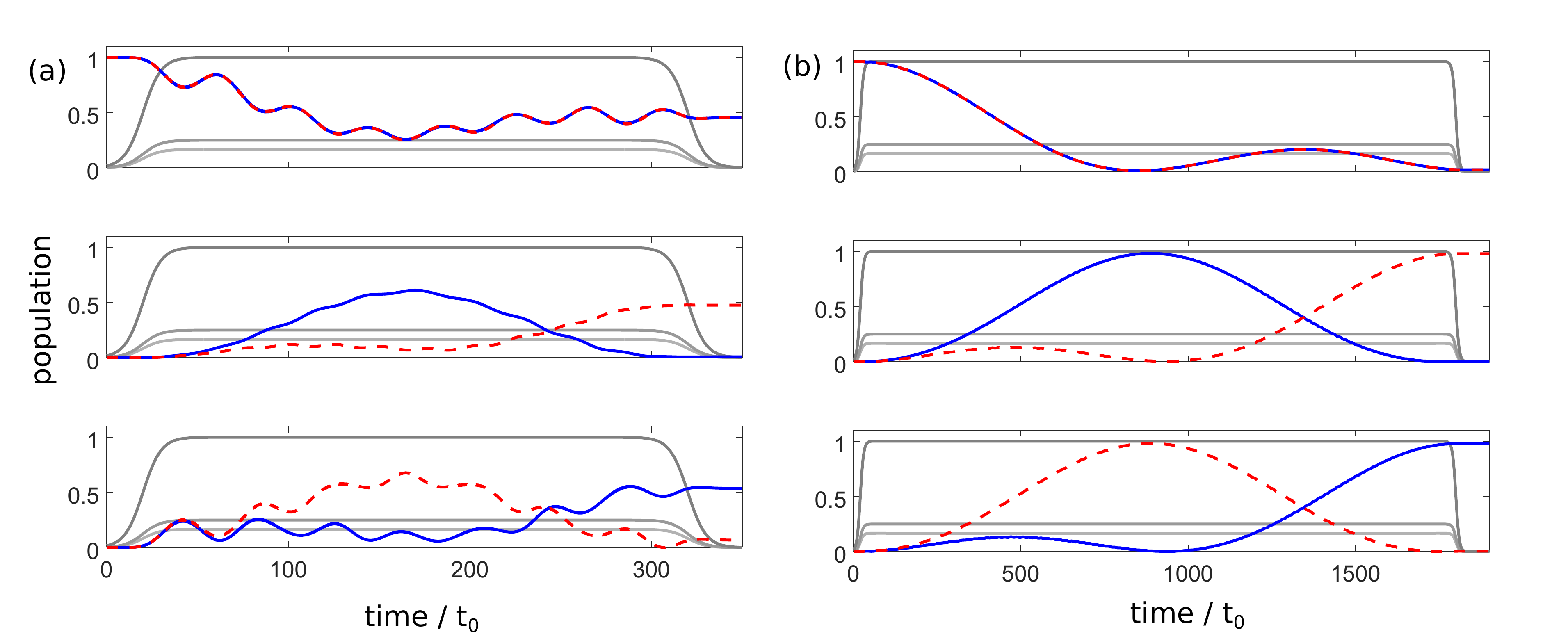}
\caption{Population of the rotational states $0_{00}$ (top),  $1_{10}$ (middle) and $1_{11}$ (bottom) of carvone.  The frequencies of the three pulses are
	 resonant to the transitions between the three states. The two enantiomers are indicated by solid (blue) and dashed (red) lines. The envelope of the pulses is indicated by gray lines.
The maximal intensities are (a) $I_z=0.14$ W/cm$^2$,  $I_y=0.38$ W/cm$^2$, and $I_x=5.64$ W/cm$^2$, (b) $I_z=0.009$ W/cm$^2$,  $I_y=0.02$ W/cm$^2$, and $I_x=0.23$ W/cm$^2$,  and time is given in units of  $t_0=\hbar/E_{rot} \approx 1.5$ ns Here, $\Phi=\pi/2$.}
\label{carvone_nonadiabatic}
\end{figure}

Non-adiabatic excitation for another chiral molecule, namely carvone (see Table~\ref{rot_constants} for the parameters) is shown in Fig.~\ref{carvone_nonadiabatic}. 
The same rotational levels are addressed as for menthol, cf. Figs.~\ref{spectrum_menthone} and \ref{fig_menthone_nonadiabatic}, but  the energy difference between the levels $1_{10}$ and $1_{11}$ is even smaller, $0.12\, B= 80.9\,$MHz in carvone compared to $0.17B=119.3 \,$MHz in methole. 
For similar field strengths as in Fig. \ref{fig_menthone_nonadiabatic} (a), only 
50 \% selectivity is obtained for carvone, cf. Fig.~\ref{carvone_nonadiabatic} (a).
In this case, for carvone, the level spacing between $1_{10}$ and $1_{11}$ is so small that the two states cannot be addressed separately. The fast oscillations of the populations in levels
$0_{00}$ and $1_{11}$ indicate additional non-resonant excitation.  By decreasing the field strength cf. Fig.~\ref{carvone_nonadiabatic} (b), non-resonant excitation is reduced and the selectivity increases. Note that here, the lowest field intensity must not exceed $I=0.009$ W/cm$^2$. 

In summary, enantio-selectivity can be realized via both adiabatic-diabatic passage and non-adiabatic excitation in real molecules such as menthol and carvone. However, care needs to be taken to identify three-level cycles with sufficient level spacing to allow for separate addressing. This becomes more difficult for heavier molecules with dense rotational spectrum. In contrast, at well separated energy levels, the interaction time can be significantly reduced by using commercially available high-power microwave radiation sources with field intensities of several W/cm$^2$.
For non-adiabatic excitation, since the fields can be applied sequentially, the two control parameters, pulse duration and field strength, can be optimized separately for each pulse.
In recent microwave  experiments  \cite{PerezAngewandte17}, enantio-selective excitation has been realized with sequences 
of partially overlapping microwave pulses. While it might have practical advantages to use overlapping pulses, 
our simulations show that for non-adiabatic excitation, temporal overlap of the pulses is not
required to achieve enantio-selectivity, as it has also been demonstrated in earlier microwave experiments \cite{LobsigerJPCL15,DomingosAnnuRevPhysChem18}.

\subsection{Application II: Ro-vibrational spectroscopy}
\label{subsec:rovib} 

A much larger frequency range can be made accessible for enantio-selective population transfer
by a combination of infrared and microwave excitation. By partly replacing the non-vanishing Cartesian projection of the permanent dipole moment by transition dipole moments, the results obtained in Sections II and III can be directly transferred to ro-vibrational excitation.  
Ro-vibrational enantio-selective excitation can thus also be applied to chiral molecules with $C_2$-symmetry which have only one non-vanishing Cartesian projection of the permanent dipole moment.
\begin{table}[tbp]
\begin{tabular}{|c|c|c|c|c|}
\hline
Transition & Type & $\mu$ [D] \cite{YurchenkoJMolSpectr2009} & Intensity [W/cm$^2$] & $ \mu E/B$ \\
       \hline
IR            & $ a$ & 0.052             & 1300      & 0.0017                        \\
IR            & $ c$ & 0.055             & 1000      & 0.0016                        \\
MW            & $b$  & 0.698             &  13       & 0.0023                       \\
\hline
\multicolumn{5}{|c|}{Rotational constants [cm$^{-1}$]}  \\
\hline
      & & $\nu=0$ \cite{WinnewisserJChemPhys2003} & $\nu(OH)=1$ \cite{BaumJMolSpectr2008} & \\
\hline      
      &A&6.740298127(45)&	6.655692(25) & \\
      &B&0.5097512033(41)&	0.5090182(16) & \\
      &C& 0.4950163369(40)&	0.4947817(17) &  \\
 \hline      
\end{tabular}
\caption{Electric dipole- and transition-dipole moments for excitation of the OH-stretch mode in HSOH and rotational constants of HSOH.
}
\label{parameter}
\end{table}
We examine enantio-selective population transfer of ro-vibrational states for HSOH as an example axially chiral molecule. Although HSOH has C$_1$-symmetry, only one component of the dipole moment is reasonably strong, namely $\mu_b$, see Table \ref{parameter}. The results can thus be directly transferred to chiral molecules with C$_2$-symmetry, e.g. HSSH. Assuming that coupling between vibrational and rotational states under field-free conditions can be neglected, the molecular eigenstates can be written as a direct product,
\begin{equation}
    |\phi_{rot} \rangle |\phi_{vib} \rangle = |J_{K_a,K_c,M} \rangle |\phi_\nu \rangle\,.
\end{equation}
We restrict our considerations to one vibrational degree of freedom, namely the OH-stretch mode of HSOH.
Note that the components of the dipole moment in the molecule-fixed coordinate system, $\hat\mu_i$ with $i=a,b,c$, are functions of the internal coordinates of the molecule, i.e., of the OH-stretch mode.
Starting in  $|0_{000}  \rangle |\phi_0 \rangle$,  
i.e., the rovibrational  ground state of HSOH,
the states $|1_{000}  \rangle |\phi_1 \rangle$ and  $|1_{10\pm1}  \rangle |\phi_1 \rangle$ are excited by $z$- and $y$-polarized infrared laser pulses, whereas  the transition between those states is driven by microwave radiation. The relevant dipole and transition dipole moments are listed in Table~\ref{parameter}. 
Since the intensity of an infrared pulse can easily be made about 100 times larger than that of microwave radiation, one can compensate for the different magnitudes of the dipole and transition dipole moments and ensure that all Rabi frequencies have approximately the same magnitude. Thus, the combination of IR and microwave radiation allows  enantio-selective excitation for classes of molecules for which pure microwave three-wave mixing spectroscopy is not possible because they have only one non-vanishing or sufficiently strong component of the permanent dipole moment.

\begin{figure}[tbp]
\centering
\includegraphics[width=16cm]{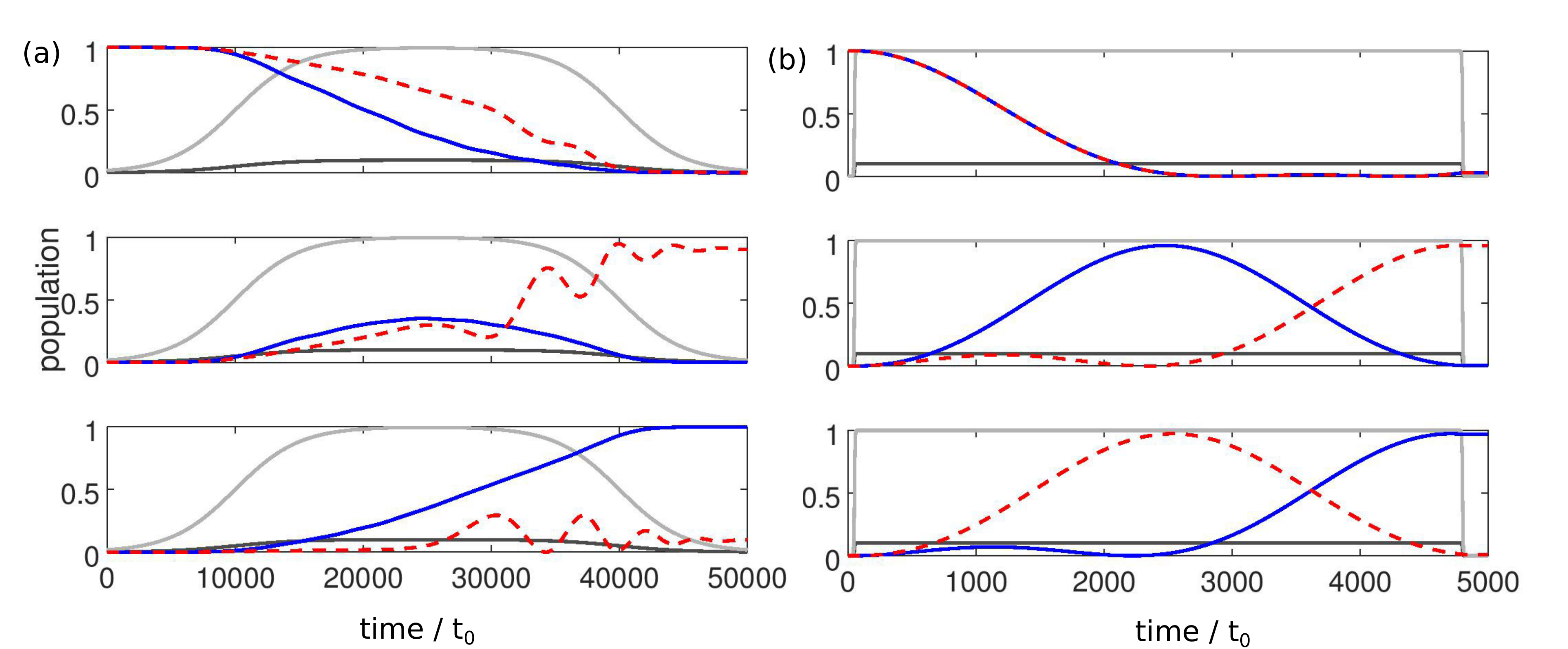}
\caption{Population transfer in HSOH for the rotational states $|\varphi_0\rangle |0_{000}\rangle$ (top),  $|\varphi_1\rangle |1_{010} \rangle$ (middle) and $|\varphi_1\rangle|1_{10\pm1}\rangle$ (bottom).  (a) Adiabatic excitation with chirped pulses for $\Phi=0$. 
(b) Non-adiabatic transfer with resonant pulses and $\Phi=\pi/2$. The two enantiomers are indicated by solid and dashed curves, respectively, and the envelopes of the pulses by gray lines (with maximal intensities denoted in Table \ref{parameter}), and time is given in units of $t_0=\hbar/E_{rot}=65\,$ps.}
\label{population_HSOH}
\end{figure}
Cyclic population transfer for HSOH in the adiabatic-diabatic (a) and non-adiabatic (b) regimes is examined in Fig.~\ref{population_HSOH}.
In Fig.~\ref{population_HSOH}(a), we observe approximately 90\% selectivity, although oscillations in the population of states  $|\varphi_1\rangle |1_{010} \rangle$ and  $|\varphi_1\rangle|1_{10\pm1}\rangle$ indicate that the process is not entirely adiabatic. 
Non-adiabatic excitation,  as shown in Fig.~\ref{population_HSOH}(b), leads to almost 100\% selectivity. These simulations suggest that enantio-selective population  transfer can also be realized by a combination of infrared and microwave pulses. Note that the pulse duration  which is necessary to obtain approximately adiabatic conditions is ten times larger than for non-adiabatic excitation. 
The sub-Doppler linewidth of infrared transitions in HSOH is of the order
of 10-100 kHz. This corresponds to lifetimes between 1.4 and 14
microseconds which need to be compared to the transfer time for the
enantio-selective population transfer of 0.3 and 3 microseconds under
non-adiabatic and adiabatic conditions, respectively. These lifetimes are
sufficiently long such that decoherence due to spontaneous emission will
not impede enantio-selective population transfer.
 
Using two infrared pulses and one microwave pulse (2IR+MW) can be regarded as an excitation scheme that shifts the critical step of population transfer to an excited vibrational
level. This is of advantage in comparison to excitation with three microwave pulses (3MW), where incoherent thermal population of rotational levels in the vibrational ground state reduces the efficacy of enantio-selective population
transfer. (2IR+MW) addresses the thermally unoccupied rotational levels in
the excited vibrational state, thus enabling coupling between fully coherent states. Any three-wave mixing scheme needs three mutually
phase coherent signals for enantio-selective population transfer. Fulfilling this condition is more involved for a (2IR + MW) scheme than for (3MW) excitation, but possible. For example, it can be realized by phase-locking two infrared
signals to a common reference standard (e.g. frequency comb) which is
controlled by a microwave reference signal. 

Enantio-selective excitation of axially chiral molecules using only infrared laser pulses and coherent excitation of three different vibrational states (3IR) 
has been studied for the example of HSSH~\cite{ThanopulosJCP03,KralPRL03,ThanopulosCPL04}. The alternative excitation (2IR+MW) scheme suggested here offers
several advantages compared to the (3IR) scheme. First, it can be  applied to
molecules which have hybrid vibrational bands with two different
dipole components. In case of HSOH, the $\nu_6$ symmetric OH-stretch vibration
can be excited by (2IR)-pulses using the weak a- and c-type vibrational
transition moments in combination with a MW-pulse driving the strong
permanent b-type transition in the excited vibrational state.
Second, compared to
(3IR)-excitation schemes, use of at least one strong permanent dipole
moment is of advantage since it increases the yield of population transfer
significantly.

\section{Conclusions}
\label{sec:concl}

We have revisited the problem of exciting the two enantiomers in a statistical mixture of randomly oriented chiral molecules to different internal states using electric dipole transitions. This is an important first step for the separation of enantiomers in gas phase experiments~\cite{EibenbergerPRL17,PerezAngewandte17}. 
Combining group theoretical considerations and numerical simulations for the simplest model as well as actual molecules, we are able to  provide a comprehensive picture of the fundamental requirements to achieve enantio-selectivitiy in different excitation regimes for 
chiral molecules with both $C_1$ and $C_2$ symmetry. 

In particular, we have provided a group-theoretical proof that population transfer within a cycle of three molecular states is enantio-selective only if electric fields with three 
mutually orthogonal polarization directions drive the transitions between the rotational  or ro-vibrational  states. This is in accordance with earlier findings \cite{LehmannJCP18} and corresponds to the experimental realization of enantio-selective three-wave mixing\cite{PattersonNature13,PattersonPCCP14,ShubertAngewandte14,LobsigerJPCL15,
	EibenbergerPRL17,PerezAngewandte17,DomingosAnnuRevPhysChem18}. It clarifies earlier misconceptions in the literature proposing excitation with only one or two different polarization directions~\cite{ShapiroPRL00,LiJCP10b,JacobJCP12,HirotaPJA12} or 
    ignoring the question of polarization directions \cite{KralPRL01,LiPRL07,LiPRA08}.
Furthermore we have identified two different regimes of enantio-selective population transfer using the three-level model: (i) Adiabatic passage of a level crossing between field-dressed eigenstates for one enantiomer with simultaneous diabatic dynamics for the other enantiomer results in enantio-selective excitation. This regime requires a specific choice of the overall phase ($\Phi=0$) and adiabatic following of the field-dressed eigenstates  which can be enforced for example by strong and linearly chirped pulses. (ii) Rabi oscillations between the rotational or ro-vibrational states lead to enantio-selective excitation for resonant non-adiabatic excitation and non-vanishing overall phase. In the latter case, the fields can be applied either simultaneously or sequentially which leaves more flexibility for pulse optimization. For example, recent experiments have used partially overlapping pulse sequences~\cite{PerezAngewandte17}, while in other experiments, non-overlapping sequences of pulses have been applied
\cite{LobsigerJPCL15,DomingosAnnuRevPhysChem18}.

For population transfer in real molecular systems, using rotational excitation  by microwave pulses in menthol and carvone and combined vibrational  and rotational excitation by infrared and microwave pulses in  HSOH, we have confirmed enantio-selectivity in both regimes but find non-adiabatic excitation to be easier to implement. It allows, in particular, for  much shorter pulse durations than the adiabatic-diabatic scenario. This is important in view of possible decoherence mechanisms. 

In the present study, we have considered molecules at $T=0$ K in order to identify fundamental limitations to enantio-selective excitation. In real experiments, even under ``cold'' conditions, many 
molecular states are thermally occupied. In particular for purely rotational excitation, practically all three levels of any conceivable cycle are initially populated. This leads to a significant reduction of selectivity~\cite{EibenbergerPRL17,PerezAngewandte17} compared to the results predicted here for zero temperature. 
Replacing at least two of the rotational transitions by transitions between two different vibrational states would allow to alleviate this problem
since, in ``cold'' experiments,  
only the lowest vibrational  state is thermally occupied, and thus
only a single state of the three-level cycle initially populated. 
This modification comes at the expense of phase locking  the infrared and microwave pulses in order to ensure a well-defined overall phase. Moreover, 
excited vibrational states have a much shorter lifetime than rotational levels in the ground vibrational manifold, introducing  spontaneous emission as a possible decoherence mechanism, in addition to collisions with other molecules.

Alternatively, one might consider to trap the molecules and cool their rotational degrees of freedom before application of the three-wave mixing spectroscopy. However, to date, cooling rotations has been limited to  diatomic molecules~\cite{VogeliusPRL02,ManaiPRL12,LienNatureComm14,ChouNature17}.
While laser cooling of polyatomic molecules might in principle be feasible~\cite{IsaevPRL16}, cooling the complex rotational structure of asymmetric top molecules is a daunting task~\cite{KochRMP}. An alternative method to prepare and measure polyatomic molecules in single quantum states has recently been proposed in Ref.~\cite{PattersonPRA2018}. Which route will eventually allow to realize full enantio-selectivity in chiral three-wave mixing experiments, currently still remains an open question.

\begin{acknowledgments}
  We would like to thank Melanie Schnell, Klaus Hornberger and Pascal Plettenberg for helpful discussions. Financial support by the Deutsche Forschungsgemeinschaft (DFG, German Research Foundation) – Projektnummer 328961117 – SFB ELCH 1319 is gratefully acknowledged.
\end{acknowledgments}

\appendix

\section{Symmetry requirements for enantio-selective excitation}
\label{sec:symmreq}
In this Appendix, we use group theoretical arguments to determine the conditions for enantio-selective excitation of rotational states. 
For this purpose, we exploit different symmetry groups. (i) The point group of a chiral molecule, C$_1$ or C$_2$, consists 
of symmetry operations that act on the internal molecular coordinates. 
It determines whether the molecule has three (C$_1$) or only one (C$_2$) non-vanishing projections 
of the permanent dipole moment. (ii) The elements of the molecular rotation group act on the Euler angles and are 
rotations about axes through the center of mass in the molecule fixed frame. For an asymmetric top, 
the molecular rotation group is $D_2$  \cite{Bunker}.
Using transformation properties with respect to the molecular rotation group, 
we prove existence of enantio-selective cycles within the molecule fixed frame
in Sec.~\ref{subsec:proof:moments}.
(iii) Rotations of an isolated molecule in free space about any axis through the center of mass in the space fixed 
coordinate system constitute the spatial rotation group K (or SO(3)). 
The resulting selection rules are expressed in form of the Wigner 3j-symbols. The spatial rotation group determines 
whether an enantio-selective cycle is independent of the spatial orientation of the molecule, i.e. whether 
it survives averaging over M-states, as we demonstrate in Sec.~\ref{subsec:proof:pol}.

\subsection{Proof that cyclic electric dipole excitation of three rotational states must contain transitions with $\mu_a$,  $\mu_b$, and $\mu_c$}
\label{subsec:proof:moments}
The symmetry group of an asymmetric top is $D_2$, its character table and the transformation properties of the asymmetric top eigenfunctions
$|J \tau M \rangle = |J_{K_a,K_c,M} \rangle$ are recalled in Table \ref{D2} \cite{Bunker}. Here $K_a$ and $K_c$ are the quantum numbers of
a prolate and oblate symmetric top, respectively.
\begin{table}
\begin{tabular}{c|cccc|c}
  $\,D_2\,$   & $\,E\,$ & $\,R_a^\pi\,$ & $\,R_b^\pi\,$ & $\,R_c^\pi\,$ & $\,K_a K_c\,$ \\
     \hline
  $A$  &   1  &   1    &   1 & 1 & ee    \\
 $B_a$   & 1  &   1    &   -1 & -1 & eo   \\
 $B_b$   & 1  &   -1    &   1 & -1 & oo   \\
  $B_c$   & 1  &   -1    &   -1 & 1 & oe   \\
 \end{tabular}
\caption{Character table of $D_2$, the molecular rotation group for asymmetric top molecules,  and transformation properties of the asymmetric top eigenfunctions~\cite{Bunker}. 
The transformation properties of the 
rotational states depend on whether the quantum numbers $K_a$ and $K_c$ are even~(e) or odd~(o).} 
\label{D2}
\end{table}
The interaction Hamiltonian, Eqs.(\ref{Hintz}), (\ref{Hintx}) and (\ref{Hinty}), can be decomposed into its irreducible parts, namely
\begin{eqnarray}
  H_{xa}, H_{ya}, H_{za} &\sim& B_a \nonumber \\
    H_{xb}, H_{yb}, H_{zb} &\sim& B_b \nonumber \\
    H_{xc}, H_{yc}, H_{zc} &\sim& B_c,
  \end{eqnarray}  
  where $H_{za}$ denotes the term of $H_{int}^z$, Eq.(\ref{Hintz}), which is proportional to $\mu_a$, and so on.
 The requirement for cyclic excitation of three molecular states to be possible at all is that all three transition matrix elements $\langle 1 | H_{int}^{(12)} | 2 \rangle$, 
 $\langle 2 | H_{int}^{(23)} | 3 \rangle$ and $\langle 3 | H_{int}^{(13)} | 1 \rangle$ must be non-zero. This can only occur if
 \begin{eqnarray}
   \Gamma (|1 \rangle) \times  \Gamma ( H_{int}^{(12)}) \times  \Gamma (| 2 \rangle) &=& A, \nonumber \\
  \Gamma (|2 \rangle) \times  \Gamma ( H_{int}^{(23)}) \times  \Gamma (| 3 \rangle) &=& A, \nonumber \\
  \Gamma (|3 \rangle) \times  \Gamma ( H_{int}^{(13)}) \times  \Gamma (|1 \rangle) &=& A,
  \label{ireps}
  \end{eqnarray}
  where $\Gamma()$ denotes the irreducible representation of the rotational states and the interaction Hamiltonians.
  Since for one-dimensional irreducible representations $\Gamma_i \times \Gamma_i =A$, Eq.(\ref{ireps}) is  fulfilled  only if
  \begin{equation}
      \Gamma ( H_{int}^{(12)}) \times  \Gamma ( H_{int}^{(23)}) \times \Gamma ( H_{int}^{(13)}) = A.
      \label{product_ireps}
  \end{equation}
  The product of the three one-dimensional  irreducible representations  can only result in the totally symmetric representation $A$
  if all three irreducible representations $B_a$, $B_b$, $B_c$ are contained in the left side of Eq.(\ref{product_ireps}). We therefore conclude that the
  three transitions have to be of $a-$type, $b-$type, and $c-$type, ind accordance with Refs.\cite{HirotaPJA12,LehmannJCP18,LiPRA18}.
  
  \subsection{Proof that enantio-selective cyclic electric dipole excitation of rotational  states requires three mutually orthogonal polarization directions}
\label{subsec:proof:pol}

\begin{figure}[tbp]
  \centering
  \includegraphics[width=14cm]{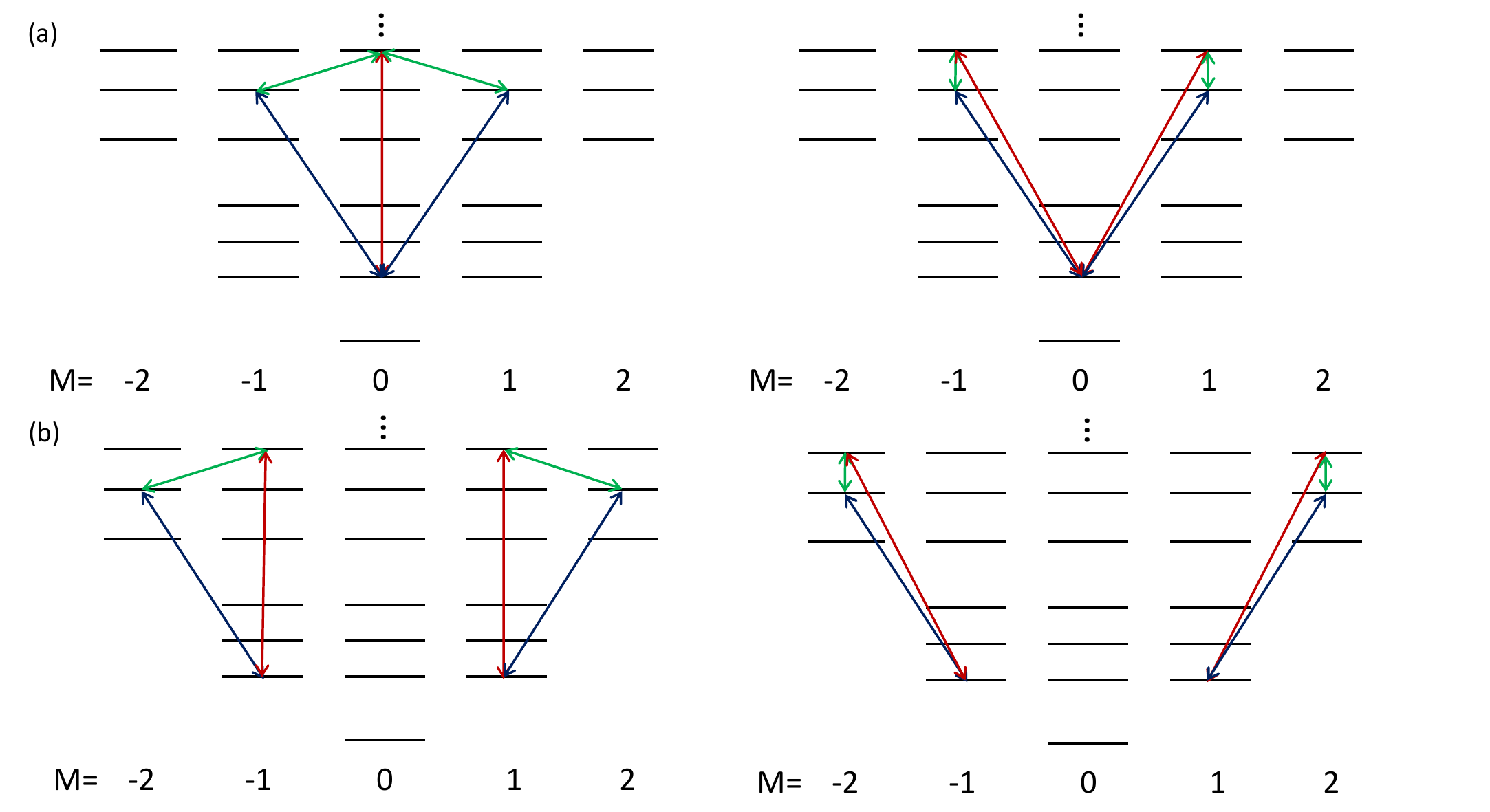}
  \caption{Cyclic excitation schemes in the presence of the $M$-degeneracy of the rotational states including states with $M=0$ (a), or with  $M\neq0$ for all states (b).}
\label{scheme2}
\end{figure}
  We consider the transition matrix elements between two rotational states of an asymmetric top, 
\begin{eqnarray}
      \langle J'', \tau'', M'' | H_{int} | J', \tau', M' \rangle &=& \sum_{K',K''} \left[ c_{K''}^{J'',M''} (\tau'') \right]^\ast c_{K'}^{J',M'} (\tau') 
\left\langle J'', K'', M'' \left| D^1_{MK} \right| J', K', M' \right\rangle,
\nonumber \\
\end{eqnarray}
where $\langle J'', K'', M'' | D^1_{MK} | J', K', M' \rangle$ are the transition matrix elements between two symmetric top eigenstates, cf. Eq.~\eqref{w3j}, 
and the coefficients  $c_{K}^{J,M} (\tau)$ are defined in  Eq.~\eqref{asym_top}.
Since transitions are allowed only between states with $\Delta J = 0, \pm1$, cyclic connection between three states must either consist of three states with the same $J$ or two states with the same $J$ and one state with $J\pm1$, see Fig.~\ref{scheme_3level}. 
    We can, furthermore, distinguish between cycles containing states with $M=0$ and cycles where $M\neq0$ for all states, as shown in Fig.~\ref{scheme2}. 
    For the latter, two equivalent cycles exist with $M>0$ and $M<0$ for all states. 
Transitions containing states with $M=0$ can be regarded as two equivalent three state cycles, sharing one or two states (those with $M=0$) \cite{LehmannJCP18}.
   We denote the three transition matrix elements of a single cycle with $H_{12}^{(\pm)}$, $H_{13}^{(\pm)}$ and  $H_{23}^{(\pm)}$,
   with the subscript $(\pm)$ referring the enantiomers.       
  An enantio-selective cycle contains one transition driven by $\mu_a$, $\mu_b$ and $\mu_c$ each, so that
  \begin{eqnarray}
      H_{cycle}^{(+)} = H_{12}^{(+)}H_{13}^{(+)}H_{23}^{(+)}=-H_{12}^{(-)}H_{13}^{(-)}H_{23}^{(-)} = -  H_{cycle}^{(-)}.
      \label{Hcycle}
     \end{eqnarray}   
 To distinguish between cycles with $M>0$ and $M<0$, we use the notation $H_{cycle}^{(\pm)} (M>0)$   and  $H_{cycle}^{(\pm)} (M<0)$, with 
 \begin{equation}
    |H_{cycle}^{(\pm)} (M>0)| =  |H_{cycle}^{(\pm)} (M<0)|.
 \end{equation}
  Cyclic excitation is enantio-selective after averaging over M-states, i.e., taking into account both equivalent cycles, if
 \begin{equation}
    H_{cycle}^{(\pm)} (M>0) = + H_{cycle}^{(\pm)} (M<0).
    \label{Hcyclep}
 \end{equation} 
Equivalently, if
\begin{equation}
    H_{cycle}^{(\pm)} (M>0) = - H_{cycle}^{(\pm)} (M<0),
    \label{Hcyclem}
 \end{equation} 
 averaging over the M-states leads to cancellation of enantio-selective effects. In order to identify enantio-selective cycles that survive M-averaging, we make use of the permutation symmetry of the Wigner 3j-symbols, which determine the transition matrix elements~\eqref{w3j}. For fields with linear polarization along the z-axis
 ($M=0$), the transition matrix elements are proportional to
 \begin{eqnarray}
    \left(    \begin{array}{ccc}  J' & 1& J' \\ M' & 0 & -M' \end{array} \right) = -  \left(    \begin{array}{ccc}  J' & 1& J' \\ -M' & 0 & +M' \end{array} \right)
    \end{eqnarray}
    for transitions with  $\Delta J = 0$ and
 \begin{eqnarray}
    \left(    \begin{array}{ccc}  J'+1 & 1& J' \\ M' & 0 & -M' \end{array} \right) =  \left(    \begin{array}{ccc}  J'+1 & 1& J' \\ -M' & 0 & +M' \end{array} \right)
    \end{eqnarray}
    for transitions with $\Delta J = \pm1$. 
    For fields with linear polarization along the x- and y-directions, $M=\pm 1$ and
    \begin{eqnarray}
    \left(    \begin{array}{ccc}  J' & 1& J' \\ M' & \pm1 & -(M'\pm1) \end{array} \right) = -  \left(    \begin{array}{ccc}  J' & 1& J' \\ -M' & \mp 1 & M'\pm1 \end{array} \right)
    \end{eqnarray}
  for $\Delta J = 0$ and
 \begin{eqnarray}
 \left(    \begin{array}{ccc}  J'+1 & 1& J' \\ M' & \pm1 & -(M'\pm1) \end{array} \right) =   \left(   
       \begin{array}{ccc}  J'+1 & 1& J' \\ -M' & \mp 1 & M'\pm1 \end{array} \right).
   \end{eqnarray}
 for $\Delta J = \pm 1$. Note further, that for x- and y-polarized fields, the interaction Hamiltonians, Eqs.~\eqref{Hintx} and \eqref{Hinty},  can be split into two parts,
        \begin{equation}
          H_{int}^{\alpha} = H_{int}^{\alpha}(M=1) + H_{int}^{\alpha}(M=-1),
          \end{equation}
          where $\alpha=x,y$ and $M=\pm1$ refers to the value of $M$ in the Wigner D-matrices $D_{MK}^J$, with
        \begin{equation}
              H_{int}^{x}(M=1) = - H_{int}^{x}(M=-1)
         \end{equation}
         and
            \begin{equation}
              H_{int}^{y}(M=1) = + H_{int}^{y} (M=-1).
         \end{equation}   
Combining all those symmetry rules, we define
\begin{equation}
     H_{ij}^{\pm} (M>0) = \sigma  H_{ij}^{\pm} (M<0)
\end{equation}
and summarize the possible values for $\sigma$ in Table~\ref{sign_transitions}. 
\begin{table}[tb]
\begin{tabular}{c|cc}
     Pol. & $\;\Delta J =0\;$ & $\;\Delta J =\pm 1\;$ \\
     \hline
  $z$  &          -1            &       +1     \\
 $x$   &            +1         &        -1     \\
 $y$   &           -1           &       +1      
 \end{tabular}
\caption{Sign $\sigma$ for transitions with $\Delta J = 0$ and $\Delta J = \pm1$ and interaction with electric fields of linear polarization along the space-fixed $z$, $x$, and $y$ axes.}
\label{sign_transitions}
\end{table}
According to Eqs.~\eqref{Hcycle} and \eqref{Hcyclep}, enantio-selective cycles which are robust to M-averaging must contain an even number of transitions with $\sigma=-1$. 
Recalling that a cycle can consist of either three transitions with $\Delta J =0$ or two transitions with $\Delta J = \pm 1$ and
one transition with $\Delta J =0$, we  conclude:
\begin{enumerate}\renewcommand{\theenumi}{\alph{enumi}}
\item Cyclic excitation with three fields with the same polarization (z-polarization) contains one or three transitions with $\sigma=-1$ and is, therefore, not enantio-selective.
\item   Cyclic excitation with two polarization directions can be realized with z-,x-,x-polarized fields or with  z-,y-,y-polarized fields.  Combinations of those transitions also lead to one or three transitions with $\sigma=-1$ and are not enantio-selective. 
\item The only way to realize enantio-selective excitation that is robust with respect to M-averaging is by using three fields with x-, y-, and z-polarization. In this case, we have either zero or two transitions with $\sigma=-1$, and thus Eq.(\ref{Hcyclep}) is fulfilled.
\end{enumerate}


\begin{thebibliography}{48}%
	\makeatletter
	\providecommand \@ifxundefined [1]{%
		\@ifx{#1\undefined}
	}%
	\providecommand \@ifnum [1]{%
		\ifnum #1\expandafter \@firstoftwo
		\else \expandafter \@secondoftwo
		\fi
	}%
	\providecommand \@ifx [1]{%
		\ifx #1\expandafter \@firstoftwo
		\else \expandafter \@secondoftwo
		\fi
	}%
	\providecommand \natexlab [1]{#1}%
	\providecommand \enquote  [1]{``#1''}%
	\providecommand \bibnamefont  [1]{#1}%
	\providecommand \bibfnamefont [1]{#1}%
	\providecommand \citenamefont [1]{#1}%
	\providecommand \href@noop [0]{\@secondoftwo}%
	\providecommand \href [0]{\begingroup \@sanitize@url \@href}%
	\providecommand \@href[1]{\@@startlink{#1}\@@href}%
	\providecommand \@@href[1]{\endgroup#1\@@endlink}%
	\providecommand \@sanitize@url [0]{\catcode `\\12\catcode `\$12\catcode
		`\&12\catcode `\#12\catcode `\^12\catcode `\_12\catcode `\%12\relax}%
	\providecommand \@@startlink[1]{}%
	\providecommand \@@endlink[0]{}%
	\providecommand \url  [0]{\begingroup\@sanitize@url \@url }%
	\providecommand \@url [1]{\endgroup\@href {#1}{\urlprefix }}%
	\providecommand \urlprefix  [0]{URL }%
	\providecommand \Eprint [0]{\href }%
	\providecommand \doibase [0]{http://dx.doi.org/}%
	\providecommand \selectlanguage [0]{\@gobble}%
	\providecommand \bibinfo  [0]{\@secondoftwo}%
	\providecommand \bibfield  [0]{\@secondoftwo}%
	\providecommand \translation [1]{[#1]}%
	\providecommand \BibitemOpen [0]{}%
	\providecommand \bibitemStop [0]{}%
	\providecommand \bibitemNoStop [0]{.\EOS\space}%
	\providecommand \EOS [0]{\spacefactor3000\relax}%
	\providecommand \BibitemShut  [1]{\csname bibitem#1\endcsname}%
	\let\auto@bib@innerbib\@empty
	\bibitem [{\citenamefont {Fujimura}\ \emph {et~al.}(2000)\citenamefont
		{Fujimura}, \citenamefont {Gonz\'alez}, \citenamefont {Hoki}, \citenamefont
		{Kr\"oner}, \citenamefont {Manz},\ and\ \citenamefont
		{Ohtsuki}}]{FujimuraAngewandte00}%
	\BibitemOpen
	\bibfield  {author} {\bibinfo {author} {\bibfnamefont {Y.}~\bibnamefont
			{Fujimura}}, \bibinfo {author} {\bibfnamefont {L.}~\bibnamefont
			{Gonz\'alez}}, \bibinfo {author} {\bibfnamefont {K.}~\bibnamefont {Hoki}},
		\bibinfo {author} {\bibfnamefont {D.}~\bibnamefont {Kr\"oner}}, \bibinfo
		{author} {\bibfnamefont {J.}~\bibnamefont {Manz}}, \ and\ \bibinfo {author}
		{\bibfnamefont {Y.}~\bibnamefont {Ohtsuki}},\ }\href {\doibase
		10.1002/1521-3773(20001215)39:24<4586::AID-ANIE4586>3.0.CO;2-H} {\bibfield
		{journal} {\bibinfo  {journal} {Angew. Chem. Int. Ed.}\ }\textbf {\bibinfo
			{volume} {39}},\ \bibinfo {pages} {4586} (\bibinfo {year}
		{2000})}\BibitemShut {NoStop}%
	\bibitem [{\citenamefont {Umeda}\ \emph {et~al.}(2002)\citenamefont {Umeda},
		\citenamefont {Takagi}, \citenamefont {Yamada}, \citenamefont {Koseki},\ and\
		\citenamefont {Fujimura}}]{UmedaJACS00}%
	\BibitemOpen
	\bibfield  {author} {\bibinfo {author} {\bibfnamefont {H.}~\bibnamefont
			{Umeda}}, \bibinfo {author} {\bibfnamefont {M.}~\bibnamefont {Takagi}},
		\bibinfo {author} {\bibfnamefont {S.}~\bibnamefont {Yamada}}, \bibinfo
		{author} {\bibfnamefont {S.}~\bibnamefont {Koseki}}, \ and\ \bibinfo {author}
		{\bibfnamefont {Y.}~\bibnamefont {Fujimura}},\ }\href {\doibase
		10.1021/ja017849f} {\bibfield  {journal} {\bibinfo  {journal} {JACS}\
		}\textbf {\bibinfo {volume} {124}},\ \bibinfo {pages} {9265} (\bibinfo {year}
		{2002})}\BibitemShut {NoStop}%
	\bibitem [{\citenamefont {Hoki}\ \emph {et~al.}(2001)\citenamefont {Hoki},
		\citenamefont {Kr\"oner},\ and\ \citenamefont {Manz}}]{HokiChemPhys01}%
	\BibitemOpen
	\bibfield  {author} {\bibinfo {author} {\bibfnamefont {K.}~\bibnamefont
			{Hoki}}, \bibinfo {author} {\bibfnamefont {D.}~\bibnamefont {Kr\"oner}}, \
		and\ \bibinfo {author} {\bibfnamefont {J.}~\bibnamefont {Manz}},\ }\href
	{\doibase http://dx.doi.org/10.1016/S0301-0104(01)00264-6} {\bibfield
		{journal} {\bibinfo  {journal} {Chem. Phys}\ }\textbf {\bibinfo {volume}
			{267}},\ \bibinfo {pages} {59 } (\bibinfo {year} {2001})}\BibitemShut
	{NoStop}%
	\bibitem [{\citenamefont {Gonz\'alez}\ \emph {et~al.}(2005)\citenamefont
		{Gonz\'alez}, \citenamefont {Manz}, \citenamefont {Schmidt},\ and\
		\citenamefont {Shibl}}]{GonzalezPCCP05}%
	\BibitemOpen
	\bibfield  {author} {\bibinfo {author} {\bibfnamefont {L.}~\bibnamefont
			{Gonz\'alez}}, \bibinfo {author} {\bibfnamefont {J.}~\bibnamefont {Manz}},
		\bibinfo {author} {\bibfnamefont {B.}~\bibnamefont {Schmidt}}, \ and\
		\bibinfo {author} {\bibfnamefont {M.~F.}\ \bibnamefont {Shibl}},\ }\href
	{\doibase 10.1039/B511495K} {\bibfield  {journal} {\bibinfo  {journal}
			{PCCP}\ }\textbf {\bibinfo {volume} {7}},\ \bibinfo {pages} {4096} (\bibinfo
		{year} {2005})}\BibitemShut {NoStop}%
	\bibitem [{\citenamefont {Shapiro}\ \emph {et~al.}(2000)\citenamefont
		{Shapiro}, \citenamefont {Frishman},\ and\ \citenamefont
		{Brumer}}]{ShapiroPRL00}%
	\BibitemOpen
	\bibfield  {author} {\bibinfo {author} {\bibfnamefont {M.}~\bibnamefont
			{Shapiro}}, \bibinfo {author} {\bibfnamefont {E.}~\bibnamefont {Frishman}}, \
		and\ \bibinfo {author} {\bibfnamefont {P.}~\bibnamefont {Brumer}},\ }\href
	{\doibase 10.1103/PhysRevLett.84.1669} {\bibfield  {journal} {\bibinfo
			{journal} {Phys. Rev. Lett.}\ }\textbf {\bibinfo {volume} {84}},\ \bibinfo
		{pages} {1669} (\bibinfo {year} {2000})}\BibitemShut {NoStop}%
	\bibitem [{\citenamefont {Kr\'al}\ and\ \citenamefont
		{Shapiro}(2001)}]{KralPRL01}%
	\BibitemOpen
	\bibfield  {author} {\bibinfo {author} {\bibfnamefont {P.}~\bibnamefont
			{Kr\'al}}\ and\ \bibinfo {author} {\bibfnamefont {M.}~\bibnamefont
			{Shapiro}},\ }\href {\doibase 10.1103/PhysRevLett.87.183002} {\bibfield
		{journal} {\bibinfo  {journal} {Phys. Rev. Lett.}\ }\textbf {\bibinfo
			{volume} {87}},\ \bibinfo {pages} {183002} (\bibinfo {year}
		{2001})}\BibitemShut {NoStop}%
	\bibitem [{\citenamefont {Kr\'al}\ \emph {et~al.}(2003)\citenamefont {Kr\'al},
		\citenamefont {Thanopulos}, \citenamefont {Shapiro},\ and\ \citenamefont
		{Cohen}}]{KralPRL03}%
	\BibitemOpen
	\bibfield  {author} {\bibinfo {author} {\bibfnamefont {P.}~\bibnamefont
			{Kr\'al}}, \bibinfo {author} {\bibfnamefont {I.}~\bibnamefont {Thanopulos}},
		\bibinfo {author} {\bibfnamefont {M.}~\bibnamefont {Shapiro}}, \ and\
		\bibinfo {author} {\bibfnamefont {D.}~\bibnamefont {Cohen}},\ }\href
	{\doibase 10.1103/PhysRevLett.90.033001} {\bibfield  {journal} {\bibinfo
			{journal} {Phys. Rev. Lett.}\ }\textbf {\bibinfo {volume} {90}},\ \bibinfo
		{pages} {033001} (\bibinfo {year} {2003})}\BibitemShut {NoStop}%
	\bibitem [{\citenamefont {Frishman}\ \emph {et~al.}(2003)\citenamefont
		{Frishman}, \citenamefont {Shapiro}, \citenamefont {Gerbasi},\ and\
		\citenamefont {Brumer}}]{FrishmanJCP03}%
	\BibitemOpen
	\bibfield  {author} {\bibinfo {author} {\bibfnamefont {E.}~\bibnamefont
			{Frishman}}, \bibinfo {author} {\bibfnamefont {M.}~\bibnamefont {Shapiro}},
		\bibinfo {author} {\bibfnamefont {D.}~\bibnamefont {Gerbasi}}, \ and\
		\bibinfo {author} {\bibfnamefont {P.}~\bibnamefont {Brumer}},\ }\href
	{\doibase {10.1063/1.1603732}} {\bibfield  {journal} {\bibinfo  {journal} {J.
				Chem. Phys.}\ }\textbf {\bibinfo {volume} {119}},\ \bibinfo {pages} {7237}
		(\bibinfo {year} {2003})}\BibitemShut {NoStop}%
	\bibitem [{\citenamefont {Thanopulos}\ \emph {et~al.}(2003)\citenamefont
		{Thanopulos}, \citenamefont {Kr\'al},\ and\ \citenamefont
		{Shapiro}}]{ThanopulosJCP03}%
	\BibitemOpen
	\bibfield  {author} {\bibinfo {author} {\bibfnamefont {I.}~\bibnamefont
			{Thanopulos}}, \bibinfo {author} {\bibfnamefont {P.}~\bibnamefont {Kr\'al}},
		\ and\ \bibinfo {author} {\bibfnamefont {M.}~\bibnamefont {Shapiro}},\ }\href
	{\doibase 10.1063/1.1597491} {\bibfield  {journal} {\bibinfo  {journal} {J.
				Chem. Phys.}\ }\textbf {\bibinfo {volume} {119}},\ \bibinfo {pages} {5105}
		(\bibinfo {year} {2003})}\BibitemShut {NoStop}%
	\bibitem [{\citenamefont {Gerbasi}\ \emph {et~al.}(2004)\citenamefont
		{Gerbasi}, \citenamefont {Brumer}, \citenamefont {Thanopulos}, \citenamefont
		{Kr\'al},\ and\ \citenamefont {Shapiro}}]{GerbasiJCP04}%
	\BibitemOpen
	\bibfield  {author} {\bibinfo {author} {\bibfnamefont {D.}~\bibnamefont
			{Gerbasi}}, \bibinfo {author} {\bibfnamefont {P.}~\bibnamefont {Brumer}},
		\bibinfo {author} {\bibfnamefont {I.}~\bibnamefont {Thanopulos}}, \bibinfo
		{author} {\bibfnamefont {P.}~\bibnamefont {Kr\'al}}, \ and\ \bibinfo {author}
		{\bibfnamefont {M.}~\bibnamefont {Shapiro}},\ }\href {\doibase
		10.1063/1.1753552} {\bibfield  {journal} {\bibinfo  {journal} {J. Chem.
				Phys.}\ }\textbf {\bibinfo {volume} {120}},\ \bibinfo {pages} {11557}
		(\bibinfo {year} {2004})}\BibitemShut {NoStop}%
	\bibitem [{\citenamefont {Jacob}\ and\ \citenamefont
		{Hornberger}(2012)}]{JacobJCP12}%
	\BibitemOpen
	\bibfield  {author} {\bibinfo {author} {\bibfnamefont {A.}~\bibnamefont
			{Jacob}}\ and\ \bibinfo {author} {\bibfnamefont {K.}~\bibnamefont
			{Hornberger}},\ }\href {\doibase 10.1063/1.4738753} {\bibfield  {journal}
		{\bibinfo  {journal} {J. Chem. Phys.}\ }\textbf {\bibinfo {volume} {137}},\
		\bibinfo {pages} {044313} (\bibinfo {year} {2012})}\BibitemShut {NoStop}%
	\bibitem [{\citenamefont {Hirota}(2012)}]{HirotaPJA12}%
	\BibitemOpen
	\bibfield  {author} {\bibinfo {author} {\bibfnamefont {E.}~\bibnamefont
			{Hirota}},\ }\href {\doibase 10.2183/pjab.88.120} {\bibfield  {journal}
		{\bibinfo  {journal} {Proc. Jpn. Acad., Ser. B}\ }\textbf {\bibinfo {volume}
			{88}},\ \bibinfo {pages} {120} (\bibinfo {year} {2012})}\BibitemShut
	{NoStop}%
	\bibitem [{\citenamefont {Lehmann}(2018)}]{LehmannJCP18}%
	\BibitemOpen
	\bibfield  {author} {\bibinfo {author} {\bibfnamefont {K.~K.}\ \bibnamefont
			{Lehmann}},\ }\href {\doibase 10.1063/1.5045052} {\bibfield  {journal}
		{\bibinfo  {journal} {J. Chem. Phys.}\ }\textbf {\bibinfo {volume} {149}},\
		\bibinfo {pages} {094201} (\bibinfo {year} {2018})}\BibitemShut {NoStop}%
	\bibitem [{\citenamefont {Vitanov}\ and\ \citenamefont
		{Drewsen}(2019)}]{Vitanov19}%
	\BibitemOpen
	\bibfield  {author} {\bibinfo {author} {\bibfnamefont {N.~V.}\ \bibnamefont
			{Vitanov}}\ and\ \bibinfo {author} {\bibfnamefont {M.}~\bibnamefont
			{Drewsen}},\ }\href@noop {} {\bibfield  {journal} {\bibinfo  {journal} {PRL}\
		} (\bibinfo {year} {2019})}\BibitemShut {NoStop}%
	\bibitem [{\citenamefont {Quack}(1986)}]{QuackCPL86}%
	\BibitemOpen
	\bibfield  {author} {\bibinfo {author} {\bibfnamefont {M.}~\bibnamefont
			{Quack}},\ }\href {\doibase 10.1016/0009-2614(86)80098-7} {\bibfield
		{journal} {\bibinfo  {journal} {Chem. Phys. Lett.}\ }\textbf {\bibinfo
			{volume} {132}},\ \bibinfo {pages} {147} (\bibinfo {year}
		{1986})}\BibitemShut {NoStop}%
	\bibitem [{\citenamefont {F\'abri}\ \emph {et~al.}(2015)\citenamefont
		{F\'abri}, \citenamefont {Horn\'y},\ and\ \citenamefont
		{Quack}}]{FabriCPC15}%
	\BibitemOpen
	\bibfield  {author} {\bibinfo {author} {\bibfnamefont {C.}~\bibnamefont
			{F\'abri}}, \bibinfo {author} {\bibfnamefont {L.}~\bibnamefont {Horn\'y}}, \
		and\ \bibinfo {author} {\bibfnamefont {M.}~\bibnamefont {Quack}},\ }\href
	{\doibase 10.1002/cphc.201500801} {\bibfield  {journal} {\bibinfo  {journal}
			{ChemPhysChem}\ }\textbf {\bibinfo {volume} {16}},\ \bibinfo {pages} {3584}
		(\bibinfo {year} {2015})}\BibitemShut {NoStop}%
	\bibitem [{\citenamefont {Patterson}\ \emph {et~al.}(2013)\citenamefont
		{Patterson}, \citenamefont {Schnell},\ and\ \citenamefont
		{Doyle}}]{PattersonNature13}%
	\BibitemOpen
	\bibfield  {author} {\bibinfo {author} {\bibfnamefont {D.}~\bibnamefont
			{Patterson}}, \bibinfo {author} {\bibfnamefont {M.}~\bibnamefont {Schnell}},
		\ and\ \bibinfo {author} {\bibfnamefont {J.~M.}\ \bibnamefont {Doyle}},\
	}\href {\doibase doi:10.1038/nature12150} {\bibfield  {journal} {\bibinfo
			{journal} {Nature}\ }\textbf {\bibinfo {volume} {497}},\ \bibinfo {pages}
		{475} (\bibinfo {year} {2013})}\BibitemShut {NoStop}%
	\bibitem [{\citenamefont {Patterson}\ and\ \citenamefont
		{Schnell}(2014)}]{PattersonPCCP14}%
	\BibitemOpen
	\bibfield  {author} {\bibinfo {author} {\bibfnamefont {D.}~\bibnamefont
			{Patterson}}\ and\ \bibinfo {author} {\bibfnamefont {M.}~\bibnamefont
			{Schnell}},\ }\href {\doibase 10.1039/C4CP00417E} {\bibfield  {journal}
		{\bibinfo  {journal} {Phys. Chem. Chem. Phys.}\ }\textbf {\bibinfo {volume}
			{16}},\ \bibinfo {pages} {11114} (\bibinfo {year} {2014})}\BibitemShut
	{NoStop}%
	\bibitem [{\citenamefont {Shubert}\ \emph {et~al.}(2014)\citenamefont
		{Shubert}, \citenamefont {Schmitz}, \citenamefont {Patterson}, \citenamefont
		{Doyle},\ and\ \citenamefont {Schnell}}]{ShubertAngewandte14}%
	\BibitemOpen
	\bibfield  {author} {\bibinfo {author} {\bibfnamefont {V.~A.}\ \bibnamefont
			{Shubert}}, \bibinfo {author} {\bibfnamefont {D.}~\bibnamefont {Schmitz}},
		\bibinfo {author} {\bibfnamefont {D.}~\bibnamefont {Patterson}}, \bibinfo
		{author} {\bibfnamefont {J.~M.}\ \bibnamefont {Doyle}}, \ and\ \bibinfo
		{author} {\bibfnamefont {M.}~\bibnamefont {Schnell}},\ }\href {\doibase
		10.1002/anie.201306271} {\bibfield  {journal} {\bibinfo  {journal} {Angew.
				Chem. Int. Ed.}\ }\textbf {\bibinfo {volume} {53}},\ \bibinfo {pages} {1152}
		(\bibinfo {year} {2014})}\BibitemShut {NoStop}%
	\bibitem [{\citenamefont {Lobsiger}\ \emph {et~al.}(2015)\citenamefont
		{Lobsiger}, \citenamefont {P\'erez}, \citenamefont {Evangelisti},
		\citenamefont {Lehmann},\ and\ \citenamefont {Pate}}]{LobsigerJPCL15}%
	\BibitemOpen
	\bibfield  {author} {\bibinfo {author} {\bibfnamefont {S.}~\bibnamefont
			{Lobsiger}}, \bibinfo {author} {\bibfnamefont {C.}~\bibnamefont {P\'erez}},
		\bibinfo {author} {\bibfnamefont {L.}~\bibnamefont {Evangelisti}}, \bibinfo
		{author} {\bibfnamefont {K.~K.}\ \bibnamefont {Lehmann}}, \ and\ \bibinfo
		{author} {\bibfnamefont {B.~H.}\ \bibnamefont {Pate}},\ }\href {\doibase
		10.1021/jz502312t} {\bibfield  {journal} {\bibinfo  {journal} {J. Phys. Chem.
				Lett.}\ }\textbf {\bibinfo {volume} {6}},\ \bibinfo {pages} {196} (\bibinfo
		{year} {2015})}\BibitemShut {NoStop}%
	\bibitem [{\citenamefont {Eibenberger}\ \emph {et~al.}(2017)\citenamefont
		{Eibenberger}, \citenamefont {Doyle},\ and\ \citenamefont
		{Patterson}}]{EibenbergerPRL17}%
	\BibitemOpen
	\bibfield  {author} {\bibinfo {author} {\bibfnamefont {S.}~\bibnamefont
			{Eibenberger}}, \bibinfo {author} {\bibfnamefont {J.}~\bibnamefont {Doyle}},
		\ and\ \bibinfo {author} {\bibfnamefont {D.}~\bibnamefont {Patterson}},\
	}\href {\doibase 10.1103/PhysRevLett.118.123002} {\bibfield  {journal}
		{\bibinfo  {journal} {Phys. Rev. Lett.}\ }\textbf {\bibinfo {volume} {118}},\
		\bibinfo {pages} {123002} (\bibinfo {year} {2017})}\BibitemShut {NoStop}%
	\bibitem [{\citenamefont {P\'erez}\ \emph {et~al.}(2017)\citenamefont
		{P\'erez}, \citenamefont {Steber}, \citenamefont {Domingos}, \citenamefont
		{Krin}, \citenamefont {Schmitz},\ and\ \citenamefont
		{Schnell}}]{PerezAngewandte17}%
	\BibitemOpen
	\bibfield  {author} {\bibinfo {author} {\bibfnamefont {C.}~\bibnamefont
			{P\'erez}}, \bibinfo {author} {\bibfnamefont {A.~L.}\ \bibnamefont {Steber}},
		\bibinfo {author} {\bibfnamefont {S.~R.}\ \bibnamefont {Domingos}}, \bibinfo
		{author} {\bibfnamefont {A.}~\bibnamefont {Krin}}, \bibinfo {author}
		{\bibfnamefont {D.}~\bibnamefont {Schmitz}}, \ and\ \bibinfo {author}
		{\bibfnamefont {M.}~\bibnamefont {Schnell}},\ }\href {\doibase
		10.1002/anie.201704901} {\bibfield  {journal} {\bibinfo  {journal}
			{Angewandte Chemie -International Edition}\ }\textbf {\bibinfo {volume}
			{56}},\ \bibinfo {pages} {12512} (\bibinfo {year} {2017})}\BibitemShut
	{NoStop}%
	\bibitem [{\citenamefont {Domingos}\ \emph {et~al.}(2018)\citenamefont
		{Domingos}, \citenamefont {P\'erez},\ and\ \citenamefont
		{Schnell}}]{DomingosAnnuRevPhysChem18}%
	\BibitemOpen
	\bibfield  {author} {\bibinfo {author} {\bibfnamefont {S.~R.}\ \bibnamefont
			{Domingos}}, \bibinfo {author} {\bibfnamefont {C.}~\bibnamefont {P\'erez}}, \
		and\ \bibinfo {author} {\bibfnamefont {M.}~\bibnamefont {Schnell}},\ }\href
	{\doibase 10.1146/annurev-physchem-052516-050629} {\bibfield  {journal}
		{\bibinfo  {journal} {Annual Review of Physical Chemistry}\ }\textbf
		{\bibinfo {volume} {69}},\ \bibinfo {pages} {499} (\bibinfo {year}
		{2018})}\BibitemShut {NoStop}%
	\bibitem [{\citenamefont {Shapiro}\ \emph {et~al.}(2003)\citenamefont
		{Shapiro}, \citenamefont {Frishman},\ and\ \citenamefont
		{Brumer}}]{ShapiroPRL03Erratum}%
	\BibitemOpen
	\bibfield  {author} {\bibinfo {author} {\bibfnamefont {M.}~\bibnamefont
			{Shapiro}}, \bibinfo {author} {\bibfnamefont {E.}~\bibnamefont {Frishman}}, \
		and\ \bibinfo {author} {\bibfnamefont {P.}~\bibnamefont {Brumer}},\ }\href
	{\doibase 10.1103/PhysRevLett.91.129902} {\bibfield  {journal} {\bibinfo
			{journal} {Phys. Rev. Lett.}\ }\textbf {\bibinfo {volume} {91}},\ \bibinfo
		{pages} {129902} (\bibinfo {year} {2003})}\BibitemShut {NoStop}%
	\bibitem [{\citenamefont {Yachmenev}\ and\ \citenamefont
		{Yurchenko}(2016)}]{YachmenevPRL16}%
	\BibitemOpen
	\bibfield  {author} {\bibinfo {author} {\bibfnamefont {A.}~\bibnamefont
			{Yachmenev}}\ and\ \bibinfo {author} {\bibfnamefont {S.~N.}\ \bibnamefont
			{Yurchenko}},\ }\href {\doibase 10.1103/PhysRevLett.117.033001} {\bibfield
		{journal} {\bibinfo  {journal} {Phys. Rev. Lett.}\ }\textbf {\bibinfo
			{volume} {117}},\ \bibinfo {pages} {033001} (\bibinfo {year}
		{2016})}\BibitemShut {NoStop}%
	\bibitem [{\citenamefont {Gershnabel}\ and\ \citenamefont
		{Averbukh}(2018)}]{GershnabelPRL18}%
	\BibitemOpen
	\bibfield  {author} {\bibinfo {author} {\bibfnamefont {E.}~\bibnamefont
			{Gershnabel}}\ and\ \bibinfo {author} {\bibfnamefont {I.~S.}\ \bibnamefont
			{Averbukh}},\ }\href {\doibase 10.1103/PhysRevLett.120.083204} {\bibfield
		{journal} {\bibinfo  {journal} {Phys. Rev. Lett.}\ }\textbf {\bibinfo
			{volume} {120}},\ \bibinfo {pages} {083204} (\bibinfo {year}
		{2018})}\BibitemShut {NoStop}%
	\bibitem [{\citenamefont {Tutunnikov}\ \emph {et~al.}(2018)\citenamefont
		{Tutunnikov}, \citenamefont {Gershnabel}, \citenamefont {Gold},\ and\
		\citenamefont {Averbukh}}]{TutunnikovJPCL18}%
	\BibitemOpen
	\bibfield  {author} {\bibinfo {author} {\bibfnamefont {I.}~\bibnamefont
			{Tutunnikov}}, \bibinfo {author} {\bibfnamefont {E.}~\bibnamefont
			{Gershnabel}}, \bibinfo {author} {\bibfnamefont {S.}~\bibnamefont {Gold}}, \
		and\ \bibinfo {author} {\bibfnamefont {I.~S.}\ \bibnamefont {Averbukh}},\
	}\href {\doibase 10.1021/acs.jpclett.7b03416} {\bibfield  {journal} {\bibinfo
			{journal} {J. Phys. Chem. Lett.}\ }\textbf {\bibinfo {volume} {9}},\
		\bibinfo {pages} {1105} (\bibinfo {year} {2018})}\BibitemShut {NoStop}%
	\bibitem [{\citenamefont {Ye}\ \emph {et~al.}(2018)\citenamefont {Ye},
		\citenamefont {Zhang},\ and\ \citenamefont {Li}}]{LiPRA18}%
	\BibitemOpen
	\bibfield  {author} {\bibinfo {author} {\bibfnamefont {C.}~\bibnamefont
			{Ye}}, \bibinfo {author} {\bibfnamefont {Q.}~\bibnamefont {Zhang}}, \ and\
		\bibinfo {author} {\bibfnamefont {Y.}~\bibnamefont {Li}},\ }\href {\doibase
		10.1103/PhysRevA.98.063401} {\bibfield  {journal} {\bibinfo  {journal} {Phys.
				Rev. A}\ }\textbf {\bibinfo {volume} {98}},\ \bibinfo {pages} {063401}
		(\bibinfo {year} {2018})}\BibitemShut {NoStop}%
	\bibitem [{\citenamefont {Ordonez}\ and\ \citenamefont
		{Smirnova}(2018)}]{OrdonezPRA18}%
	\BibitemOpen
	\bibfield  {author} {\bibinfo {author} {\bibfnamefont {A.~F.}\ \bibnamefont
			{Ordonez}}\ and\ \bibinfo {author} {\bibfnamefont {O.}~\bibnamefont
			{Smirnova}},\ }\href {\doibase 10.1103/PhysRevA.98.063428} {\bibfield
		{journal} {\bibinfo  {journal} {Phys. Rev. A}\ }\textbf {\bibinfo {volume}
			{98}},\ \bibinfo {pages} {063428} (\bibinfo {year} {2018})}\BibitemShut
	{NoStop}%
	\bibitem [{\citenamefont {Zare}(1988)}]{Zare88}%
	\BibitemOpen
	\bibfield  {author} {\bibinfo {author} {\bibfnamefont {R.~N.}\ \bibnamefont
			{Zare}},\ }\href@noop {} {\emph {\bibinfo {title} {Angular Momentum}}}\
	(\bibinfo  {publisher} {Wiley},\ \bibinfo {year} {1988})\BibitemShut
	{NoStop}%
	\bibitem [{\citenamefont {Vitanov}\ \emph {et~al.}(2001)\citenamefont
		{Vitanov}, \citenamefont {Halfmann}, \citenamefont {Shore},\ and\
		\citenamefont {Bergmann}}]{VitanovAnuRevPhysChem01}%
	\BibitemOpen
	\bibfield  {author} {\bibinfo {author} {\bibfnamefont {N.~V.}\ \bibnamefont
			{Vitanov}}, \bibinfo {author} {\bibfnamefont {T.}~\bibnamefont {Halfmann}},
		\bibinfo {author} {\bibfnamefont {B.~W.}\ \bibnamefont {Shore}}, \ and\
		\bibinfo {author} {\bibfnamefont {K.}~\bibnamefont {Bergmann}},\ }\href
	{\doibase 10.1146/annurev.physchem.52.1.763} {\bibfield  {journal} {\bibinfo
			{journal} {Annual Review of Physical Chemistry}\ }\textbf {\bibinfo {volume}
			{52}},\ \bibinfo {pages} {763} (\bibinfo {year} {2001})}\BibitemShut
	{NoStop}%
	\bibitem [{\citenamefont {Schmitz}\ \emph {et~al.}(2015)\citenamefont
		{Schmitz}, \citenamefont {Shubert}, \citenamefont {Betz},\ and\ \citenamefont
		{Schnell}}]{SchmitzFrontiers15}%
	\BibitemOpen
	\bibfield  {author} {\bibinfo {author} {\bibfnamefont {D.}~\bibnamefont
			{Schmitz}}, \bibinfo {author} {\bibfnamefont {V.~A.}\ \bibnamefont
			{Shubert}}, \bibinfo {author} {\bibfnamefont {T.}~\bibnamefont {Betz}}, \
		and\ \bibinfo {author} {\bibfnamefont {M.}~\bibnamefont {Schnell}},\ }\href
	{\doibase 10.3389/fchem.2015.00015} {\bibfield  {journal} {\bibinfo
			{journal} {Frontiers in Chemistry}\ }\textbf {\bibinfo {volume} {3}},\
		\bibinfo {pages} {15} (\bibinfo {year} {2015})}\BibitemShut {NoStop}%
	\bibitem [{\citenamefont {Moreno}\ \emph {et~al.}(2013)\citenamefont {Moreno},
		\citenamefont {Huet},\ and\ \citenamefont
		{Gonz{\'a}lez}}]{MorenoStructural13}%
	\BibitemOpen
	\bibfield  {author} {\bibinfo {author} {\bibfnamefont {J.~R.~A.}\
			\bibnamefont {Moreno}}, \bibinfo {author} {\bibfnamefont {T.~R.}\
			\bibnamefont {Huet}}, \ and\ \bibinfo {author} {\bibfnamefont {J.~J.~L.}\
			\bibnamefont {Gonz{\'a}lez}},\ }\href {\doibase 10.1007/s11224-012-0142-8}
	{\bibfield  {journal} {\bibinfo  {journal} {Structural Chemistry}\ }\textbf
		{\bibinfo {volume} {24}},\ \bibinfo {pages} {1163} (\bibinfo {year}
		{2013})}\BibitemShut {NoStop}%
	\bibitem [{\citenamefont {Yurchenko}\ \emph {et~al.}(2009)\citenamefont
		{Yurchenko}, \citenamefont {Yachmenev}, \citenamefont {Thiel}, \citenamefont
		{Baum}, \citenamefont {Giesen}, \citenamefont {Melnikov},\ and\ \citenamefont
		{Jensen}}]{YurchenkoJMolSpectr2009}%
	\BibitemOpen
	\bibfield  {author} {\bibinfo {author} {\bibfnamefont {S.~N.}\ \bibnamefont
			{Yurchenko}}, \bibinfo {author} {\bibfnamefont {A.}~\bibnamefont
			{Yachmenev}}, \bibinfo {author} {\bibfnamefont {W.}~\bibnamefont {Thiel}},
		\bibinfo {author} {\bibfnamefont {O.}~\bibnamefont {Baum}}, \bibinfo {author}
		{\bibfnamefont {T.~F.}\ \bibnamefont {Giesen}}, \bibinfo {author}
		{\bibfnamefont {V.~V.}\ \bibnamefont {Melnikov}}, \ and\ \bibinfo {author}
		{\bibfnamefont {P.}~\bibnamefont {Jensen}},\ }\href@noop {} {\bibfield
		{journal} {\bibinfo  {journal} {J. Mol. Spectrosc.}\ }\textbf {\bibinfo
			{volume} {257}},\ \bibinfo {pages} {57} (\bibinfo {year} {2009})}\BibitemShut
	{NoStop}%
	\bibitem [{\citenamefont {Winnewisser}\ \emph {et~al.}(2003)\citenamefont
		{Winnewisser}, \citenamefont {Lewen}, \citenamefont {Thorwirth},
		\citenamefont {Behnke}, \citenamefont {Hahn}, \citenamefont {J.Gauss},\ and\
		\citenamefont {Herbst}}]{WinnewisserJChemPhys2003}%
	\BibitemOpen
	\bibfield  {author} {\bibinfo {author} {\bibfnamefont {G.}~\bibnamefont
			{Winnewisser}}, \bibinfo {author} {\bibfnamefont {F.}~\bibnamefont {Lewen}},
		\bibinfo {author} {\bibfnamefont {S.}~\bibnamefont {Thorwirth}}, \bibinfo
		{author} {\bibfnamefont {M.}~\bibnamefont {Behnke}}, \bibinfo {author}
		{\bibfnamefont {J.}~\bibnamefont {Hahn}}, \bibinfo {author} {\bibnamefont
			{J.Gauss}}, \ and\ \bibinfo {author} {\bibfnamefont {E.}~\bibnamefont
			{Herbst}},\ }\href@noop {} {\bibfield  {journal} {\bibinfo  {journal} {Chem.
				Eur. J.}\ }\textbf {\bibinfo {volume} {9}},\ \bibinfo {pages} {5501}
		(\bibinfo {year} {2003})}\BibitemShut {NoStop}%
	\bibitem [{\citenamefont {Baum}\ \emph {et~al.}(2008)\citenamefont {Baum},
		\citenamefont {Giesen},\ and\ \citenamefont
		{Schlemmer}}]{BaumJMolSpectr2008}%
	\BibitemOpen
	\bibfield  {author} {\bibinfo {author} {\bibfnamefont {O.}~\bibnamefont
			{Baum}}, \bibinfo {author} {\bibfnamefont {T.}~\bibnamefont {Giesen}}, \ and\
		\bibinfo {author} {\bibfnamefont {S.}~\bibnamefont {Schlemmer}},\ }\href@noop
	{} {\bibfield  {journal} {\bibinfo  {journal} {J. Mol. Spectrosc.}\ }\textbf
		{\bibinfo {volume} {247}},\ \bibinfo {pages} {25} (\bibinfo {year}
		{2008})}\BibitemShut {NoStop}%
	\bibitem [{\citenamefont {Thanopulos}\ \emph {et~al.}(2004)\citenamefont
		{Thanopulos}, \citenamefont {Paspalakis},\ and\ \citenamefont
		{Kis}}]{ThanopulosCPL04}%
	\BibitemOpen
	\bibfield  {author} {\bibinfo {author} {\bibfnamefont {I.}~\bibnamefont
			{Thanopulos}}, \bibinfo {author} {\bibfnamefont {E.}~\bibnamefont
			{Paspalakis}}, \ and\ \bibinfo {author} {\bibfnamefont {Z.}~\bibnamefont
			{Kis}},\ }\href {\doibase http://dx.doi.org/10.1016/j.cplett.2004.03.129}
	{\bibfield  {journal} {\bibinfo  {journal} {Chem. Phys. Lett.}\ }\textbf
		{\bibinfo {volume} {390}},\ \bibinfo {pages} {228} (\bibinfo {year}
		{2004})}\BibitemShut {NoStop}%
	\bibitem [{\citenamefont {Li}\ and\ \citenamefont {Shapiro}(2010)}]{LiJCP10b}%
	\BibitemOpen
	\bibfield  {author} {\bibinfo {author} {\bibfnamefont {X.}~\bibnamefont
			{Li}}\ and\ \bibinfo {author} {\bibfnamefont {M.}~\bibnamefont {Shapiro}},\
	}\href {\doibase 10.1063/1.3429884} {\bibfield  {journal} {\bibinfo
			{journal} {J. Chem. Phys.}\ }\textbf {\bibinfo {volume} {132}},\ \bibinfo
		{pages} {194315} (\bibinfo {year} {2010})}\BibitemShut {NoStop}%
	\bibitem [{\citenamefont {Li}\ \emph {et~al.}(2007)\citenamefont {Li},
		\citenamefont {Bruder},\ and\ \citenamefont {Sun}}]{LiPRL07}%
	\BibitemOpen
	\bibfield  {author} {\bibinfo {author} {\bibfnamefont {Y.}~\bibnamefont
			{Li}}, \bibinfo {author} {\bibfnamefont {C.}~\bibnamefont {Bruder}}, \ and\
		\bibinfo {author} {\bibfnamefont {C.~P.}\ \bibnamefont {Sun}},\ }\href
	{\doibase 10.1103/PhysRevLett.99.130403} {\bibfield  {journal} {\bibinfo
			{journal} {Phys. Rev. Lett.}\ }\textbf {\bibinfo {volume} {99}},\ \bibinfo
		{pages} {130403} (\bibinfo {year} {2007})}\BibitemShut {NoStop}%
	\bibitem [{\citenamefont {Li}\ and\ \citenamefont {Bruder}(2008)}]{LiPRA08}%
	\BibitemOpen
	\bibfield  {author} {\bibinfo {author} {\bibfnamefont {Y.}~\bibnamefont
			{Li}}\ and\ \bibinfo {author} {\bibfnamefont {C.}~\bibnamefont {Bruder}},\
	}\href {\doibase 10.1103/PhysRevA.77.015403} {\bibfield  {journal} {\bibinfo
			{journal} {Phys. Rev. A}\ }\textbf {\bibinfo {volume} {77}},\ \bibinfo
		{pages} {015403} (\bibinfo {year} {2008})}\BibitemShut {NoStop}%
	\bibitem [{\citenamefont {Vogelius}\ \emph {et~al.}(2002)\citenamefont
		{Vogelius}, \citenamefont {Madsen},\ and\ \citenamefont
		{Drewsen}}]{VogeliusPRL02}%
	\BibitemOpen
	\bibfield  {author} {\bibinfo {author} {\bibfnamefont {I.~S.}\ \bibnamefont
			{Vogelius}}, \bibinfo {author} {\bibfnamefont {L.~B.}\ \bibnamefont
			{Madsen}}, \ and\ \bibinfo {author} {\bibfnamefont {M.}~\bibnamefont
			{Drewsen}},\ }\href {\doibase 10.1103/PhysRevLett.89.173003} {\bibfield
		{journal} {\bibinfo  {journal} {Phys. Rev. Lett.}\ }\textbf {\bibinfo
			{volume} {89}},\ \bibinfo {pages} {173003} (\bibinfo {year}
		{2002})}\BibitemShut {NoStop}%
	\bibitem [{\citenamefont {Manai}\ \emph {et~al.}(2012)\citenamefont {Manai},
		\citenamefont {Horchani}, \citenamefont {Lignier}, \citenamefont {Pillet},
		\citenamefont {Comparat}, \citenamefont {Fioretti},\ and\ \citenamefont
		{Allegrini}}]{ManaiPRL12}%
	\BibitemOpen
	\bibfield  {author} {\bibinfo {author} {\bibfnamefont {I.}~\bibnamefont
			{Manai}}, \bibinfo {author} {\bibfnamefont {R.}~\bibnamefont {Horchani}},
		\bibinfo {author} {\bibfnamefont {H.}~\bibnamefont {Lignier}}, \bibinfo
		{author} {\bibfnamefont {P.}~\bibnamefont {Pillet}}, \bibinfo {author}
		{\bibfnamefont {D.}~\bibnamefont {Comparat}}, \bibinfo {author}
		{\bibfnamefont {A.}~\bibnamefont {Fioretti}}, \ and\ \bibinfo {author}
		{\bibfnamefont {M.}~\bibnamefont {Allegrini}},\ }\href {\doibase
		10.1103/PhysRevLett.109.183001} {\bibfield  {journal} {\bibinfo  {journal}
			{Phys. Rev. Lett.}\ }\textbf {\bibinfo {volume} {109}},\ \bibinfo {pages}
		{183001} (\bibinfo {year} {2012})}\BibitemShut {NoStop}%
	\bibitem [{\citenamefont {Lien}\ \emph {et~al.}(2014)\citenamefont {Lien},
		\citenamefont {Seck}, \citenamefont {Lin}, \citenamefont {Nguyen},
		\citenamefont {Tabor},\ and\ \citenamefont {Odom}}]{LienNatureComm14}%
	\BibitemOpen
	\bibfield  {author} {\bibinfo {author} {\bibfnamefont {C.-Y.}\ \bibnamefont
			{Lien}}, \bibinfo {author} {\bibfnamefont {C.~M.}\ \bibnamefont {Seck}},
		\bibinfo {author} {\bibfnamefont {Y.-W.}\ \bibnamefont {Lin}}, \bibinfo
		{author} {\bibfnamefont {J.~H.}\ \bibnamefont {Nguyen}}, \bibinfo {author}
		{\bibfnamefont {D.~A.}\ \bibnamefont {Tabor}}, \ and\ \bibinfo {author}
		{\bibfnamefont {B.~C.}\ \bibnamefont {Odom}},\ }\href {\doibase
		10.1038/ncomms5783} {\bibfield  {journal} {\bibinfo  {journal} {Nature
				Commun.}\ }\textbf {\bibinfo {volume} {5}},\ \bibinfo {pages} {4783}
		(\bibinfo {year} {2014})}\BibitemShut {NoStop}%
	\bibitem [{\citenamefont {Chou}\ \emph {et~al.}(2017)\citenamefont {Chou},
		\citenamefont {Kurz}, \citenamefont {Hume}, \citenamefont {Plessow},
		\citenamefont {Leibrandt},\ and\ \citenamefont {Leibfried}}]{ChouNature17}%
	\BibitemOpen
	\bibfield  {author} {\bibinfo {author} {\bibfnamefont {C.-W.}\ \bibnamefont
			{Chou}}, \bibinfo {author} {\bibfnamefont {C.}~\bibnamefont {Kurz}}, \bibinfo
		{author} {\bibfnamefont {D.~B.}\ \bibnamefont {Hume}}, \bibinfo {author}
		{\bibfnamefont {P.~N.}\ \bibnamefont {Plessow}}, \bibinfo {author}
		{\bibfnamefont {D.~R.}\ \bibnamefont {Leibrandt}}, \ and\ \bibinfo {author}
		{\bibfnamefont {D.}~\bibnamefont {Leibfried}},\ }\href {\doibase
		10.1038/nature22338} {\bibfield  {journal} {\bibinfo  {journal} {Nature}\
		}\textbf {\bibinfo {volume} {545}},\ \bibinfo {pages} {203–207} (\bibinfo
		{year} {2017})}\BibitemShut {NoStop}%
	\bibitem [{\citenamefont {Isaev}\ and\ \citenamefont
		{Berger}(2016)}]{IsaevPRL16}%
	\BibitemOpen
	\bibfield  {author} {\bibinfo {author} {\bibfnamefont {T.~A.}\ \bibnamefont
			{Isaev}}\ and\ \bibinfo {author} {\bibfnamefont {R.}~\bibnamefont {Berger}},\
	}\href {\doibase 10.1103/PhysRevLett.116.063006} {\bibfield  {journal}
		{\bibinfo  {journal} {Phys. Rev. Lett.}\ }\textbf {\bibinfo {volume} {116}},\
		\bibinfo {pages} {063006} (\bibinfo {year} {2016})}\BibitemShut {NoStop}%
	\bibitem [{\citenamefont {Koch}\ \emph {et~al.}(2018)\citenamefont {Koch},
		\citenamefont {Lemeshko},\ and\ \citenamefont {Sugny}}]{KochRMP}%
	\BibitemOpen
	\bibfield  {author} {\bibinfo {author} {\bibfnamefont {C.~P.}\ \bibnamefont
			{Koch}}, \bibinfo {author} {\bibfnamefont {M.}~\bibnamefont {Lemeshko}}, \
		and\ \bibinfo {author} {\bibfnamefont {D.}~\bibnamefont {Sugny}},\
	}\href@noop {} {\bibfield  {journal} {\bibinfo  {journal} {arXiv:1810.11338}\
		} (\bibinfo {year} {2018})}\BibitemShut {NoStop}%
	\bibitem [{\citenamefont {Patterson}(2018)}]{PattersonPRA2018}%
	\BibitemOpen
	\bibfield  {author} {\bibinfo {author} {\bibfnamefont {D.}~\bibnamefont
			{Patterson}},\ }\href@noop {} {\bibfield  {journal} {\bibinfo  {journal}
			{Phys. Rev. A}\ }\textbf {\bibinfo {volume} {97}},\ \bibinfo {pages} {033403}
		(\bibinfo {year} {2018})}\BibitemShut {NoStop}%
	\bibitem [{\citenamefont {Bunker}\ and\ \citenamefont {Jensen}(1998)}]{Bunker}%
	\BibitemOpen
	\bibfield  {author} {\bibinfo {author} {\bibfnamefont {P.~R.}\ \bibnamefont
			{Bunker}}\ and\ \bibinfo {author} {\bibfnamefont {P.}~\bibnamefont
			{Jensen}},\ }\enquote {\bibinfo {title} {Molecular symmetry and
			spectroscopy},}\ \ (\bibinfo  {publisher} {NRC Research Press},\ \bibinfo
	{year} {1998})\ Chap.\ \bibinfo {chapter} {12.6}\BibitemShut {NoStop}%
\end{thebibliography}
%

\end{document}